\documentclass{aastex701}

	\usepackage{booktabs}  % 用于更好的表格线
	\usepackage{geometry}  % 用于调整页边距
	\usepackage{longtable}
	\usepackage{amsmath}
	\usepackage{hyperref}

\begin{document}
%			\begin{CJK*}{UTF8}{gbsn}
\title{The Internal Magnetic Field Structure of ICMEs in the Heliosphere}

				\author[0009-0004-0819-0301]{Ziwei Huang}
				\affiliation{Planetary Environmental and Astrobiological Research Laboratory (PEARL), School of Atmospheric Sciences, Sun Yat-sen University, Zhuhai, China}
				\affiliation{State Key Laboratory of Solar Activity and Space Weather, National Space Science Center, Chinese Academy of Sciences, Beijing, China}
				\affiliation{CAS Center for Excellence in Comparative Planetology/CAS Key Laboratory of Geospace Environment, University of Science and Technology of China, Hefei 230026, People's Republic of China}
				\email{huangzw28@mail2.sysu.edu.cn}
				
				\author[0000-0001-7276-3208]{Zhenjun Zhou}
				\affiliation{Planetary Environmental and Astrobiological Research Laboratory (PEARL), School of Atmospheric Sciences, Sun Yat-sen University, Zhuhai, China}
				\affiliation{State Key Laboratory of Solar Activity and Space Weather, National Space Science Center, Chinese Academy of Sciences, Beijing, China}
				\affiliation{CAS Center for Excellence in Comparative Planetology/CAS Key Laboratory of Geospace Environment, University of Science and Technology of China, Hefei 230026, People's Republic of China}
				\email[show]{zhouzhj7@mail.sysu.edu.cn}  
				
				\author[0000-0002-1854-8459]{Yudong Ye}
				\affiliation{Planetary Environmental and Astrobiological Research Laboratory (PEARL), School of Atmospheric Sciences, Sun Yat-sen University, Zhuhai, China}
				\affiliation{CAS Center for Excellence in Comparative Planetology/CAS Key Laboratory of Geospace Environment, University of Science and Technology of China, Hefei 230026, People's Republic of China}
				\email{yeyd5@mail.sysu.edu.cn}  
				
				\author[0000-0001-9427-7366]{Ming Xiong}
                \affiliation{State Key Laboratory of Solar Activity and Space Weather, National Space Science Center, Chinese Academy of Sciences, Beijing, China}
                \affiliation{College of Earth and Planetary Sciences, University of Chinese Academy of Sciences, Beijing 100049, People’s Republic of China}
                \email{mxiong@swl.ac.cn}  
				
				\author[0000-0002-8887-3919]{Yuming Wang}
				\affiliation{National Key Laboratory of Deep Space Exploration/School of Earth and Space Sciences, University of Science and Technology of China, Hefei 230026, People's Republic of China}
				\affiliation{CAS Center for Excellence in Comparative Planetology/CAS Key Laboratory of Geospace Environment, University of Science and Technology of China, Hefei 230026, People's Republic of China}
				\email{ymwang@ustc.edu.cn}
				
				\author[0000-0001-9315-4487]{Yutian Chi}
				\affiliation{National Key Laboratory of Deep Space Exploration/Institute of Deep Space Sciences, Deep Space Exploration Laboratory, Hefei, 230088, Anhui, China}
				\email{ytchi@mail.ustc.edu.cn}

				\author[0000-0001-7894-8246]{Daniel Heyner}
				\affiliation{Institut f\"ur Geophysik und extraterrestrische Physik, Technische Universit\"at Braunschweig, Braunschweig, 38106, Germany}
				\email{d.heyner@tu-braunschweig.de}
				
				\author{Hans-Ulrich Auster}
				\affiliation{Institut f\"ur Geophysik und extraterrestrische Physik, Technische Universit\"at Braunschweig, Braunschweig, 38106, Germany}
				\email{uli.auster@tu-bs.de}
				
				\author[0000-0002-5324-4039]{Ingo Richter}
				\affiliation{Institut f\"ur Geophysik und extraterrestrische Physik, Technische Universit\"at Braunschweig, Braunschweig, 38106, Germany}
				\email{i.richter@tu-bs.de}
				
				\author[0000-0003-0277-3253]{Beatriz S\'anchez-Cano}
				\affiliation{School of Physics and Astronomy, University of Leicester, Leicester,
					LE1 7RH, UK}
				\email{bscmdr1@leicester.ac.uk}

				%% Use the \collaboration command to identify collaborations. This command
				%% takes an optional argument that is either a number or the word "all"
				%% which tells the compiler how many of the authors above the command to
				%% show. For example "\collaboration[all]{(DELVE Collaboration)}" wil include
				%% all the authors above this command.
				%%
				%% Mark off the abstract in the ``abstract'' environment. 

\begin{abstract}
	Interplanetary coronal mass ejections (ICMEs) are major drivers of heliospheric disturbances and space-weather effects. Here we present a multi-spacecraft study of 96 magnetic clouds (MCs) distributed over a broad range of heliocentric distances, and reconstruct their internal magnetic structure with a uniform-twist Gold--Hoyle (GH) flux-rope model. From the fits, we derive the axial field strength $B_0$, the twist density (turn density) $\tau$, the GH parameter $\omega$, and the integrated twist number $n$.
	% We find that $B_0$ and $\tau$ both decrease with increasing heliocentric distance, consistent with expansion and axial stretching during propagation. A key result is that the physically relevant characteristic quantity is the twist density $\tau$ (rather than the integrated twist number $n$): in the $\tau$--$R$ plane, the events are bounded by an upper envelope with an approximate $R^{-1}$ dependence, equivalent in GH variables to a near-constant boundary around $\omega\!\sim\!2$. This behavior implies that, for a given flux-rope scale, the winding density exhibits a well-defined empirical upper boundary. In contrast, $n$ does not exhibit a comparable, systematic upper-bound evolution with distance. 
	\added{We find that $B_0$ and the turn density $\tau$ both decrease with increasing heliocentric distance, consistent with expansion and axial stretching during propagation. A key result is that the upper envelope in the $\tau$--$R$ plane corresponds to a nearly constant boundary in the dimensionless GH parameter $\omega=2\pi R\tau$, close to $\omega\sim2$. Therefore, the inferred upper value of $\tau$ is not scale-independent, but follows $\tau_{\max}\simeq \omega_{\max}/(2\pi R)$ for a given flux-rope radius. In contrast, the estimated integrated turn number $n$ shows no similarly clear radial organization in the present sample.}
	This study investigates the ICME structure and magnetic field characteristics across heliocentric distances from 0.07 to 5.4~AU, thereby providing observational constraints on the large-scale evolution of interplanetary magnetic flux ropes.
\end{abstract}

				%% Keywords should appear after the \end{abstract} command. 
			%% The AAS Journals now uses Unified Astronomy Thesaurus (UAT) concepts:
			%% https://astrothesaurus.org
			%% You will be asked to selected these concepts during the submission process
			%% but this old "keyword" functionality is maintained in case authors want
			%% to include these concepts in their preprints.
			%%
			%% You can use the \uat command to link your UAT concepts back its source.
			
\section{Introduction}

Coronal mass ejections (CMEs) are large-scale eruptions of coherent magnetic-field structures and associated solar plasma from the corona into interplanetary space, and are among the primary drivers of severe space weather throughout the heliosphere.
Their propagation and interaction with planetary environments can produce intense
geomagnetic storms, posing risks to space-borne and ground-based technological systems
\citep{Schwenn2006,Pulkkinen2007}.
Understanding the internal magnetic structure of CMEs is therefore essential for
establishing the physical connection between coronal eruptions and their heliospheric
manifestations.

\begin{figure*}[ht!]
	\plotone{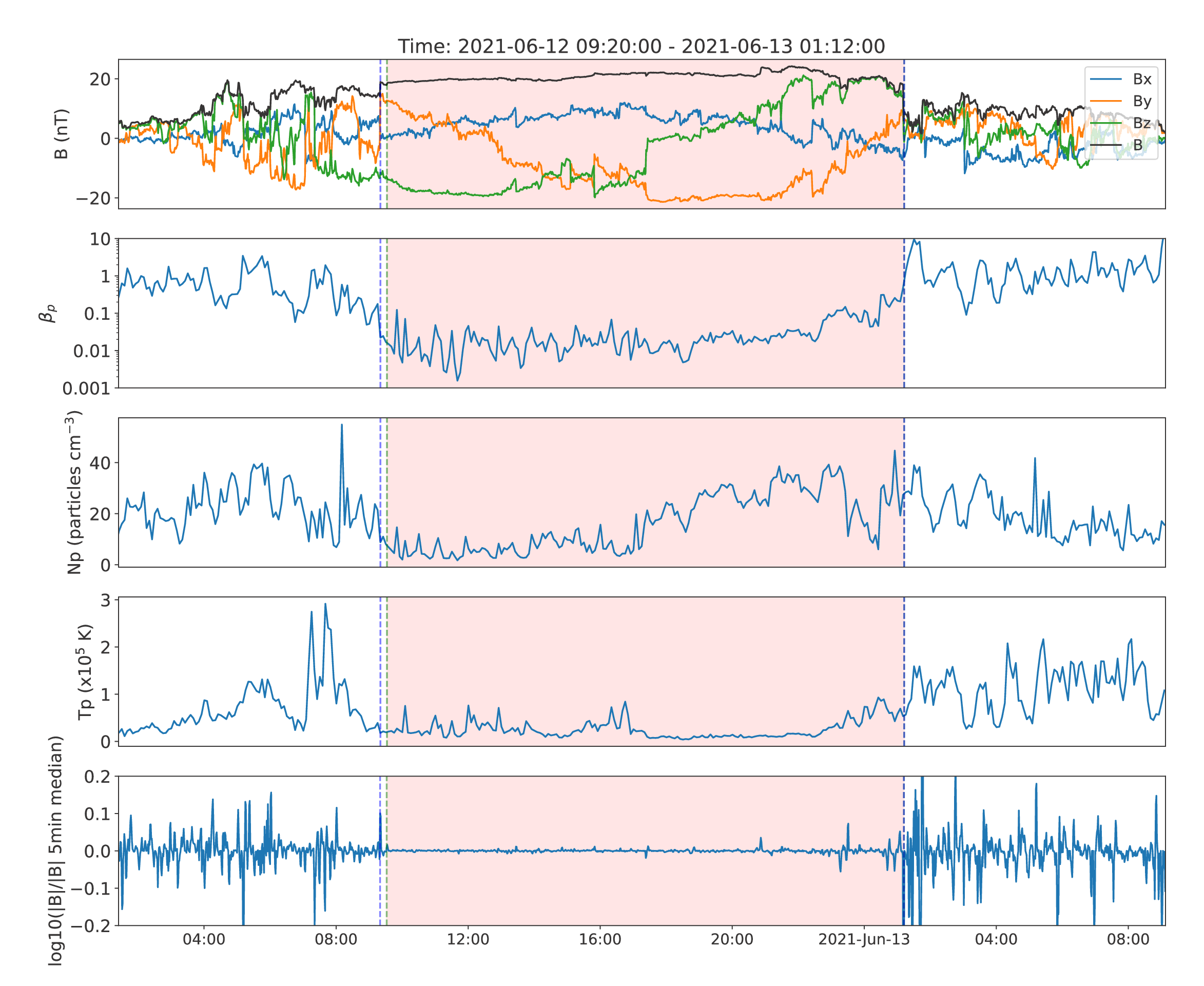}
	\caption{Overview of solar wind parameters for an MC event (No.~61 in Table~\ref{tab:long}). From top to bottom, the panels show: (1) the total magnetic field strength $|B|$ (black) and its three components ($B_x$, $B_y$, $B_z$) in RTN coordinates; (2) the plasma beta ($\beta_p$); (3) the proton number density $N_p$ (particles~cm$^{-3}$); (4) the proton temperature $T_p$ ($\times 10^5$~K); (5) $\log_{10}(|B|/\langle|B|\rangle)$, where $\langle|B|\rangle$ is the 5-minute moving average of $|B|$. The pink--shaded region marks the MC interval identified in this work, while the blue vertical dashed lines indicate the MC boundaries reported in the ICMECAT catalog.
		\label{fig:mc_obs}}
\end{figure*}

The magnetic flux rope (MFR) is widely regarded as the core magnetic structure of CMEs, with a coherent helical field geometry. Magnetic clouds (MCs; Figure~\ref{fig:mc_obs}) are a well-defined subset of ICMEs, identified by enhanced magnetic field strength, smooth large-scale field rotation, low plasma beta, and depressed proton temperature \citep{Burlaga1981, KleinBurlaga1982}. Together, these in situ signatures indicate that many ICMEs retain a flux-rope configuration during heliospheric propagation.

Remote-sensing observations further support the existence of MFRs in the solar corona,
including sigmoidal soft X-ray structures prior to eruption, helical or kinked morphologies
in coronagraph images, and strong rotational motions observed during eruptive jets \citep{Liu2014,Zhou2019,Zhou2020,Zhou2022}.
Despite differences in spatial scale, global geometry, and surrounding magnetic environment,
these phenomena consistently indicate the release of magnetic free energy stored in
twisted magnetic fields. Recent high-resolution coronal observations enabled by
solar adaptive optics have further revealed rapidly evolving, finely twisted substructures
in the low corona, providing direct observational context for the buildup and release
of magnetic twist in eruptive systems \citep{Schmidt2025,Zhou2025}.

The degree of magnetic twist within an MFR plays a central role in both its formation
and eruption.
Magnetic twist is fundamentally a geometric quantity that measures how magnetic field
lines wind around the flux-rope axis.
It is locally defined as
\begin{equation}
	T(r) = \frac{1}{r}\frac{B_\phi(r)}{B_z(r)},
\end{equation}
where $B_\phi$ and $B_z$ are the azimuthal and axial components of the magnetic field,
respectively, and $r$ is the distance from the axis.
The total twist angle accumulated along the axis of an MFR is then given by
\begin{equation}
	\Phi_T = \int_0^{l} T(r)\,dz,
\end{equation}
with $l$ denoting the axial length of the flux rope, and the corresponding number of
turns is $n = \Phi_T / 2\pi$.

Magnetic twist is closely related to magnetic helicity and free magnetic energy, and
excessive twist can render an MFR unstable.
Several theoretical models have proposed characteristic twist thresholds for magnetic flux ropes under different idealized assumptions. For example, the classical Kruskal--Shafranov criterion \citep{Shafranov1957,Kruskal1958} and the line-tied model of \citet{HoodPriest1981} describe stability conditions for force-free equilibrium configurations, while the model of \citet{DungeyLoughhead1954} emphasizes the role of the global aspect ratio through the relation $\Phi_c \approx 2l/R$. Although these models are derived under different physical assumptions, they suggest that both the total twist and the local winding density may play important roles in twisted magnetic structures. However, interplanetary magnetic clouds evolve dynamically during propagation, and the relative importance of these quantities in observational ICME structures remains unclear.

Observational estimates of pre-eruptive twist in coronal MFRs show a wide range of values.
Nonlinear force-free field extrapolations suggest relatively modest twist values, typically
of order $\sim \pi$ to a few $\pi$, for both eruptive and non-eruptive structures
\citep[e.g.,][]{Yan2001, Regnier2002, Guo2010,Inoue2011, Inoue2012}.
In contrast, multi-wavelength observations of prominences, coronal loops, and eruptive
jets often imply much higher twist values, in some cases exceeding $10\pi$
\citep[e.g.,][]{Vrsnak1991, Romano2003,  Gary2004,Srivastava2010}.
This apparent discrepancy raises a fundamental question: which estimates more faithfully
represent the true magnetic structure of erupting flux ropes?

Insight into this issue can be gained from in situ measurements of ICMEs, where the magnetic
field is directly sampled.
Using a velocity-modified uniform-twist Gold--Hoyle flux-rope model,
\citet{Wang2016} analyzed a large sample of MCs observed at 1~AU and found that most
interplanetary flux ropes are highly twisted, with the majority exceeding two full turns.
They further showed that the twist scales with the aspect ratio of the flux rope and
appears to be bounded by an upper limit.
These results suggest that strongly twisted magnetic field lines may constrain the
expansion and overall size of interplanetary flux ropes.
Complementary event-based evidence was later provided by \citet{Wang2017}, who linked
solar and in-situ observations for the same eruption and showed that substantial twist can be
built up during the flare, while the reconstructed radial twist profile in the corona is broadly
consistent with that of the associated magnetic cloud near 1~AU.
In addition, \citet{Wang2018} found that the axial magnetic flux of magnetic cloud 
declines during outward propagation, with such structural variations jointly driven 
by the pancaking effect and the erosion effect. In the heliospheric propagation of MCs, the twist 
can be comprehensively influenced by combined processes including magnetic reconnection and erosion.
Together, these studies support a coherent coronal--interplanetary observational picture of
flux-rope twist evolution.

In this study, we extend such analyses by reconstructing the magnetic structures of ICMEs
distributed at different heliocentric distances and heliolongitudes within the ecliptic
plane using the Gold--Hoyle uniform-twist flux-rope model.
By systematically estimating the magnetic twist of interplanetary MFRs and examining its
dependence on global geometry and heliocentric distance, 
we aim to quantify the twist properties and evolutionary constraints of dynamically evolving MFRs.

\begin{table}[ht]
	\caption{Model Parameters Used in the GH Flux-Rope Analysis}
	\label{tab:par}
	\small
	\centering
	\begin{tabular}{ll}
		\toprule\toprule
		\textbf{Parameter} & \textbf{Explanation} \\
		\midrule
		
		\multicolumn{2}{l}{\textit{Free Parameters in the Model}} \\[2pt]
		
		$B_0$      & Magnetic field intensity at the center of the magnetic flux rope \\
		$\omega$   & Parameter characterizing the magnetic twist of the flux rope \\
		$\theta$   & Inclination angle of the flux-rope axis in RTN coordinates \\
		$\phi$     & Azimuthal orientation of the flux-rope axis in RTN coordinates \\
		$d$        & Minimum distance between the spacecraft trajectory and the flux-rope axis \\
		
		\multicolumn{2}{l}{\textit{Quantities Derived from the Model}} \\[2pt]
		
		$R_t$     & $\Delta t$, cross-sectional scale in the time domain (hours) \\
		$t_c$      & Time at which the spacecraft reaches its closest approach to the axis \\
		$T_t$     & Twist rate along the axis, defined as $\omega / R_t$ \\
		$\tau_t$  & Number of field-line turns per unit time, given by $T_t / 2\pi$ \\
		$\chi_n$   & Normalized RMS deviation between modeled magnetic fields and in situ observations \\
		$Q$        & A quality flag used to evaluate the reliability and robustness of the fitted magnetic-flux-rope parameters \\
		\bottomrule
	\end{tabular}
\end{table}

			\section{Analysis Method and Selection of MC Events}\label{sec:method}

			\subsection{Introduction to the Gold-Hoyle (GH) Model}
			
			In this study, the GH model, originally proposed by~\cite{Gold1960}, is applied to reconstruct magnetic clouds in order to obtain their structural properties, including the axis direction (elevation and azimuthal angles), magnetic field strength at the axis of the MFR, \(\omega\), radius, and other related parameters.
			This GH model describes a force-free magnetic flux rope with uniform twist. It assumes that the magnetic field lines wrap helically around a central axis with constant pitch, making it a useful analytical model for studying MCs.
			The model is nonlinear force-free, satisfying \(\nabla \times \mathbf{B} = \alpha \mathbf{B}\), where \(\alpha\) is a spatially varying force-free parameter:
			
			\[
			\alpha(r) = \frac{2T}{1 + T^2 r^2}
			\]

			In cylindrical coordinates \((r, \phi, z)\), the magnetic field components in the GH model are given by:
			
			\[
			B_r = 0, \quad
			B_\phi = \frac{T r}{1 + T^2 r^2} B_0, \quad
			B_z = \frac{B_0}{1 + T^2 r^2}
			\]
			
			where:
			\begin{itemize}
				\item \(B_0\) is the magnetic field strength at the axis (\(r = 0\)),
				\item \(T\) is the twist per unit length,
				\item \(r\) is the radial distance from the axis.
			\end{itemize}

In the GH uniform-twist magnetic-flux-rope model, the geometric boundary of the flux rope is not uniquely prescribed, implying a certain degree of freedom in defining its cross-sectional scale. To explicitly relate the size of the flux rope to its magnetic twist, we introduce a parameter $\omega$ that connects the characteristic radius $R$ with the axial twist rate $T$ through
\begin{equation}
	R = \frac{\omega}{T}.
\end{equation}
Here, $\omega$ is treated as an event-dependent parameter in the GH model, defined by
$\omega=RT$. This quantity provides a compact way to connect rope size and local winding.
Because the total twist satisfies
$n=\Phi_T/(2\pi)=Tl/(2\pi)=(\omega/2\pi)(l/R)$,
the mapping between a critical total twist and a critical $\omega$ is model-dependent
and requires an explicit geometric assumption for the aspect ratio $l/R$
(e.g., fixed $l/R$ or an adopted $l/R$ scaling).

Within the idealized GH framework, the ratio between the flux-rope radius and axial length,
as well as the total twist angle $\Phi_T=2\pi n$, can be treated as invariants for a given
rope realization. Under this assumption, $\omega$ is also treated as approximately invariant
during propagation. We note, however, that fitted $\omega$ values are event-dependent in
observations. With this parametrization, the magnetic-field components inside the flux rope are

\[
B_r = 0, \quad
B_{\varphi} = \frac{\omega x}{1 + \omega^{2} x^{2}}\, B_0, \quad
B_{z} = \frac{1}{1 + \omega^{2} x^{2}}\, B_0,
\]

where $x=r/R$ denotes the normalized radial distance from the flux-rope axis and $B_0$
is the axial magnetic-field strength at the center.

Because reliable solar-wind speed data are unavailable for part of the MC sample, we fit the GH model to magnetic-field measurements only. We parameterize the flux-rope size in time units by assuming an approximately constant solar-wind speed within each MC interval, so that time can be used as a proxy for length. All model parameters and their physical meanings are summarized in Table~\ref{tab:par}.

\begin{figure}[htbp]
	\centering
	\includegraphics[width=\textwidth]{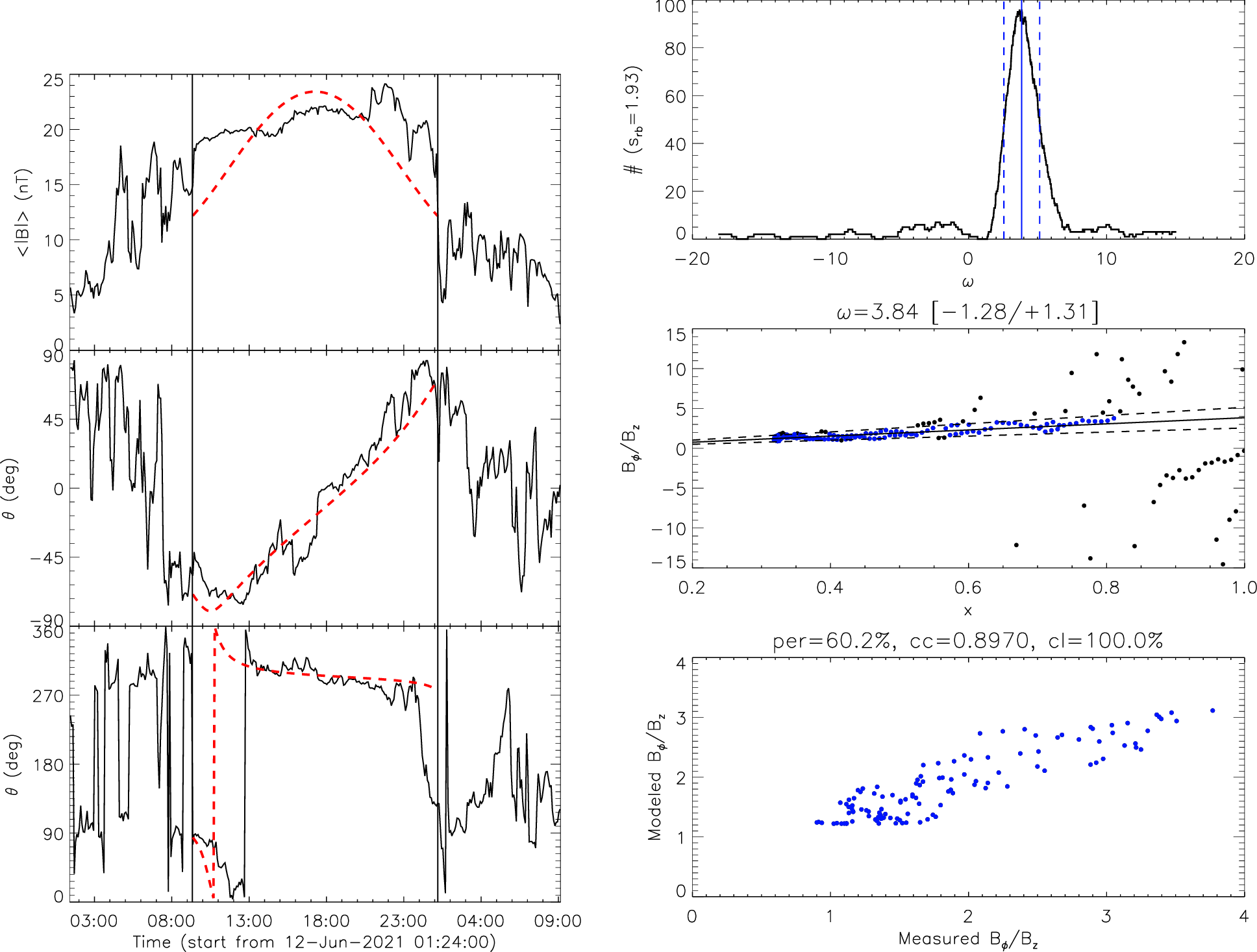}
	\caption{
		(Left) Interplanetary magnetic field recorded by the PSP spacecraft for MC No.~61 in Table~\ref{tab:long}. From top to bottom, the panels show the total magnetic field strength, the elevation angle, and the azimuthal angle of the magnetic field vector in RTN coordinates. The red dashed curves represent the GH model fitting results. 
		(Right) The top panel displays the histogram of the twist parameter $\omega$. The middle panel shows the relationship between $x$ and the ratio $\frac{B_\phi}{B_z}$ within the MC. The bottom panel presents the correlation between the modeled and observed values of $\frac{B_\phi}{B_z}$.
	}
	\label{fig:mc_fit}
\end{figure}

Compared with other models, such as the Lundquist model \citep{Lundquist1950,Wang2015} which assumes linear force-free fields and nonuniform twist, the GH model provides a simpler and often more realistic approximation for magnetic flux ropes with nearly uniform twist, particularly in interplanetary space. It is frequently used for reconstructing magnetic clouds and estimating magnetic helicity, twist angle, and energy content. Figure~\ref{fig:mc_fit} shows the GH-model reconstruction for event No.~61 in Table~\ref{tab:long} as an example.

			\begin{figure*}[ht!]
				\plotone{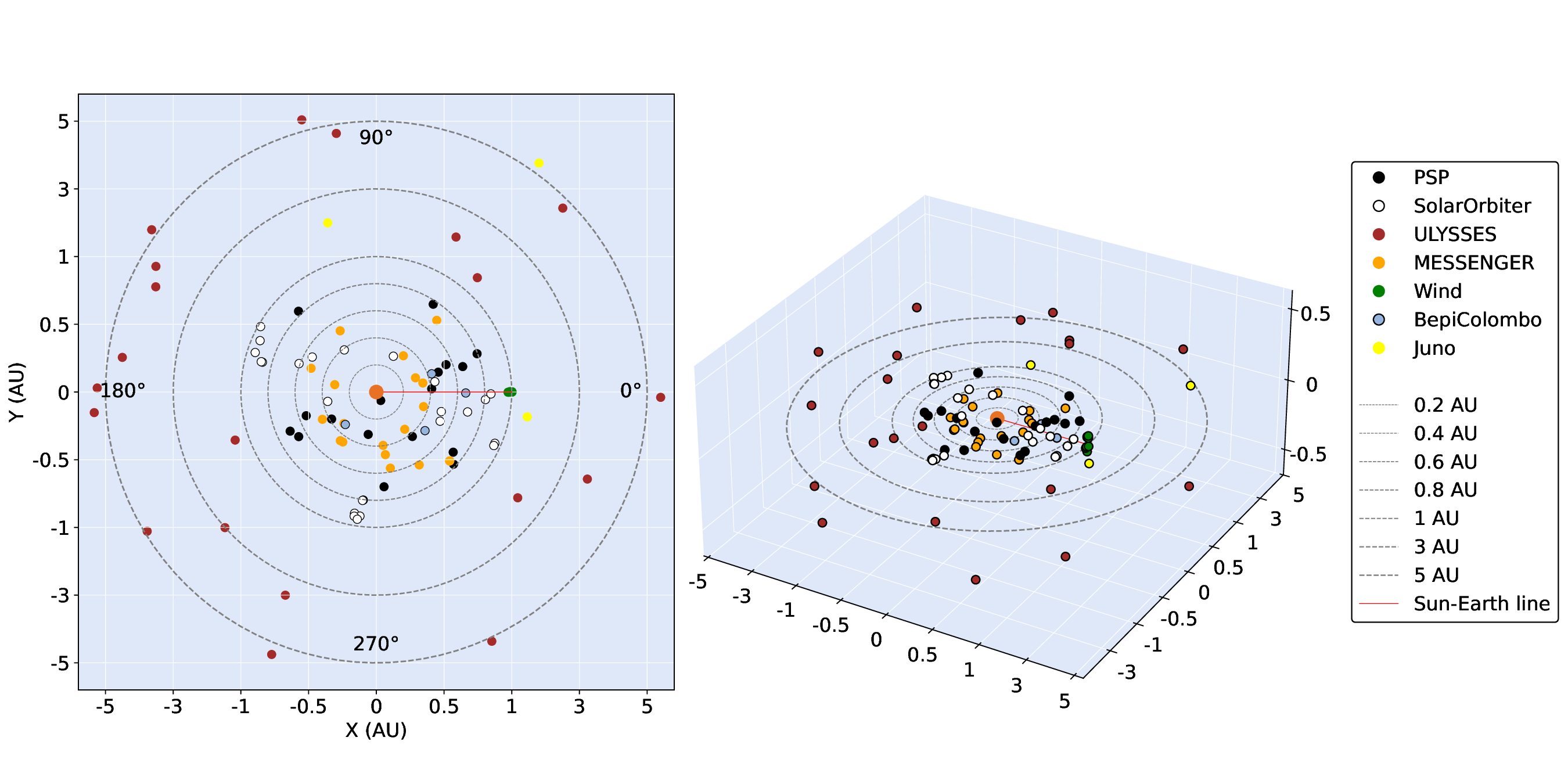}
				\caption{Positions of the selected ICMECAT events plotted in HEEQ coordinates. The left panel shows a top-down (XY plane) view, while the right panel provides a 3D perspective. The Earth (green dot) is fixed at $0^\circ$ longitude. Each marker represents an MC event observed by one of seven spacecraft. The radial distance is given in astronomical units (AU).
					\label{fig:ICMECAT_HEEQ}}
			\end{figure*}

			\subsection{Assessment of Fitting Quality for the GH Model}

			The quality of a fitting result for the GH model can be evaluated using the following steps:
			
			\subsubsection{Goodness-of-Fit Evaluation}
			The fitting procedure is carried out using a least-squares fitting method. The key metric to evaluate the fit quality is the normalized root mean square error \(\chi_n\), which is calculated as:
			
			\[
			\chi_n = \sqrt{\frac{1}{2N} \sum_{i=1}^{N} \left(\frac{B_{m,i} - B_{o,i}}{B_{o,i}}\right)^2}
			\]
			
			where the \( B_{m,i} \) and \( B_{o,i} \) are the modeled and observed magnetic field components, respectively, \( N \) is the number of data points. A value of \(\chi_n\) closer to 0 indicates a better fit.
			
			\subsubsection{Uniform-Twist Assumption Validation}
			The uniform-twist assumption of the GH model is evaluated by estimating the parameter \(\omega\) as:
			
			\[
			\omega =RT=\frac{B_\phi}{x B_z}
			\]
			
			The goodness of the uniform-twist configuration is assessed by the percentage (per) of data points that fall within the uncertainty range of \(\omega\), the correlation coefficient (cc) between the modeled and observed magnetic field components and the confidence level (cl) of the correlation under the permutation test. See the right three panels in Figure~\ref{fig:mc_fit}.

\subsubsection{Systematic Overestimation Factor for GH-based Twist}\label{subsubsec:overfactor}

A known caveat of the uniform-twist GH reconstruction is a systematic overestimation of twist-related quantities. Following \citet{Wang2016}, we adopt an empirical overestimation factor of 2.5 and apply it event by event in this work. Specifically, all GH-derived twist-related quantities are scaled down by a factor of 2.5 before statistical analysis and plotting.

The physical basis is the calibration reported by \citet{Wang2016}: the twist inferred from the velocity-modified GH model shows an almost perfect linear correlation with independently probed twist estimates (cc $\approx$ 0.9987), but with systematically larger magnitudes by about a factor of 2.5 on average (after excluding a few outliers). Therefore, we treat the 2.5 scaling as an empirical correction to the absolute twist magnitude.

Hereafter, all twist-related quantities reported in this paper (e.g., $\tau$, $\tau_t$, $\tau_{\rm AU}$, $\omega$, and $n$) are corrected values by default, unless explicitly stated otherwise. The event list table is the only exception and retains the original (uncorrected) fitted values for traceability.

\subsubsection{Selection Criteria Based on Fitting Results}
An acceptable fitting in this study is required to satisfy the following basic criteria:
\begin{enumerate}
	\item the normalized deviation parameter must fulfill $\chi_n \leq 0.5$, indicating a good agreement
	between the model and observations based on the RMS error;
	\item the normalized closest approach $d$ must be smaller than $0.7$, ensuring that the spacecraft
	trajectory passes sufficiently deep inside the magnetic cloud rather than merely skimming its outer
	boundary.
\end{enumerate}
Events that do not satisfy either of the above two criteria are excluded from further analysis.

			To further assess the reliability of the fitting, we introduce a quality flag $Q$.
			Events that additionally satisfy $\text{per} \geq 30\%$, $\text{cc} \geq 0.8$, and
			$\text{cl} \geq 80\%$ are classified as $Q = 1$, representing high-quality fittings.
			Events that meet the basic acceptance criteria but fail to satisfy at least one of these
			three additional conditions are classified as $Q = 2$.
			
			%(Möstl et al. 2017).

\subsection{Selection of MC Events}
The magnetic cloud (MC) candidates analyzed in this study were first identified from the
HELIO4CAST ICME catalog (ICMECAT; \citealt{Mostl2017,Mostl2020}), which compiles ICME
events observed by multiple spacecraft across a broad range of heliocentric distances.
From this catalog, we selected MC candidates according to the following criteria:
\begin{enumerate}
	\item Each candidate must exhibit well-defined magnetic-cloud signatures, characterized by an enhanced
	and smoothly rotating magnetic-field vector over an extended interval.
	\item The minimum MC duration is required to be longer than 3~hours, in order to exclude small-scale
	flux-rope structures embedded in the ambient solar wind.
	\item All candidates are further screened by the GH-model fitting requirements described in
	Section~\ref{sec:method} (Selection Criteria Based on Fitting Results), including the basic acceptance
	criteria and the quality-flag classification.
\end{enumerate}

\subsection{Magnetic Cloud Boundary Identification Method}

Magnetic cloud boundaries were determined by analyzing fluctuations in the magnetic-field magnitude, see the bottom panel in Fig~\ref{fig:mc_obs}.
The field strength was normalized to its 5-minute sliding mean, \(B/\langle B \rangle\), and the temporal
variation of this ratio was used as the primary boundary indicator. Based on this profile, initial boundary
candidates were manually selected, and the final boundaries were then determined by applying a threshold of
\(5\,\sigma_{\log_{10}(B/\langle B \rangle)}\), where \(\sigma\) is the standard deviation of
\(\log_{10}(B/\langle B \rangle)\) calculated over the central two-thirds of the cloud duration.
Whenever the identified boundaries deviated significantly from the expected morphology, the results were
re-examined together with plasma parameters to ensure reliability.
The physical motivation and a more detailed analysis underlying this boundary identification method
will be presented in a forthcoming study.

\section{Statistical Properties of the ICMEs}

\subsection{Axial field strength of MCs}
\begin{figure*}[ht!]
				\plotone{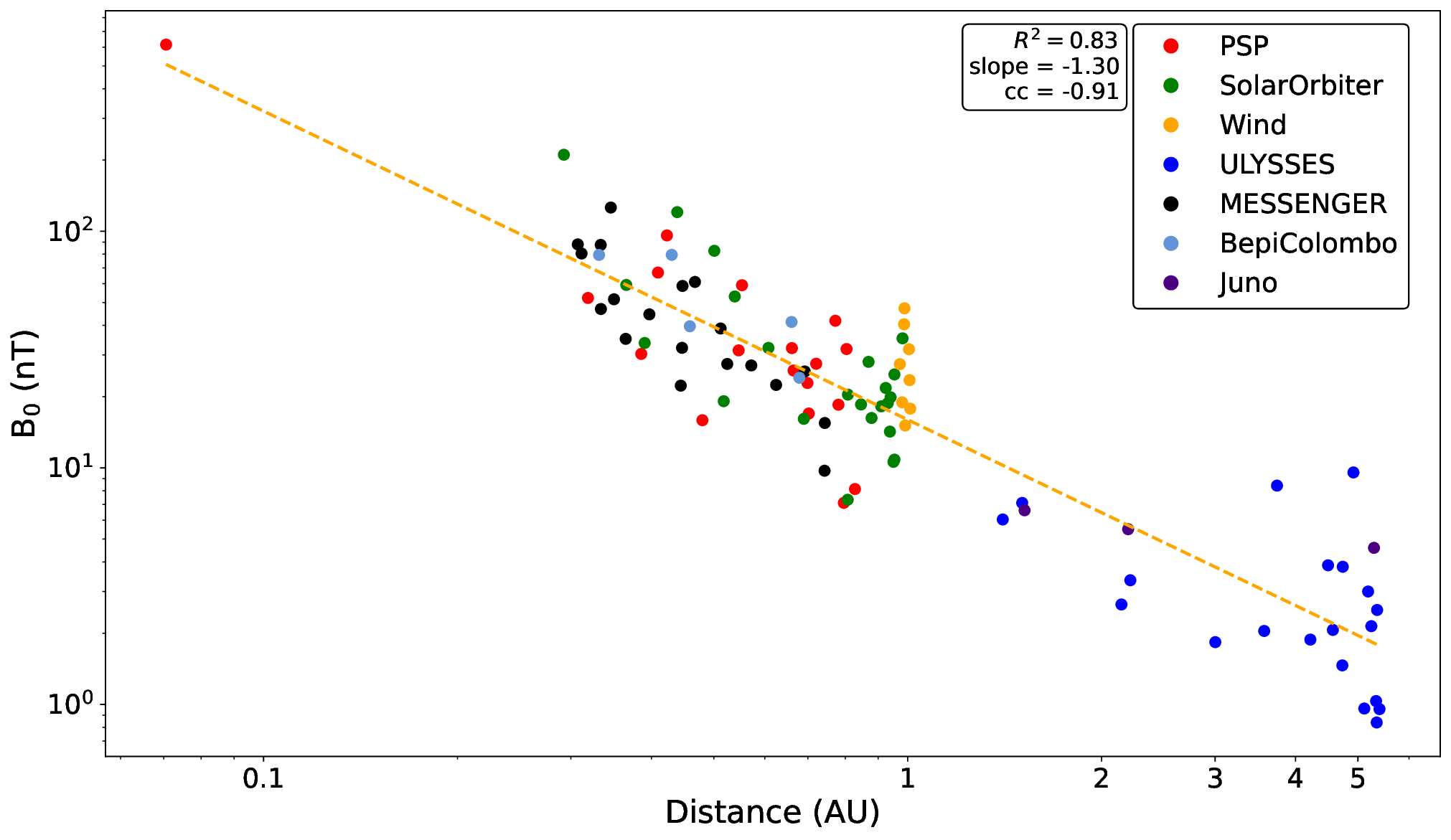}
				\caption{Magnetic field strength at the flux rope axis ($B_0$) as a function of the heliocentric distance in astronomical units (AU). Each colored dot represents a magnetic cloud event observed by a different spacecraft, as indicated in the legend. The orange dashed line denotes the best-fit power-law relation between $B_0$ and distance, with the corresponding fitting parameters ($R^2$, slope, and correlation coefficient) shown in the inset. A clear decrease of $B_0$ with increasing heliocentric distance is evident.
					\label{fig:B0_vs_r}}
\end{figure*}

Figure~\ref{fig:B0_vs_r} summarizes how the fitted axial field strength of MCs varies with heliocentric distance.  
The abscissa is the heliocentric distance on a logarithmic scale (in AU), and the ordinate is the GH--model axial field strength
$B_0$ (in nT). Each colored marker corresponds to one of the 96 MC events, with the color indicating the observing spacecraft.
A linear regression performed in log--log space (orange dashed line) yields
\begin{equation}
	\log_{10}\!\left(\frac{B_0}{\mathrm{nT}}\right)
	= a + b\,\log_{10}\!\left(\frac{r}{\mathrm{AU}}\right),
	\qquad b=-1.30,
\end{equation}
with $R^{2}=0.83$ and a correlation coefficient ${\rm cc}=-0.91$ (Figure~\ref{fig:B0_vs_r}). Equivalently, the trend can be
written as $B_0 \propto r^{-1.30}$, indicating a systematic decay of the MC core field with increasing distance.
Although the 96 events mostly correspond to independent MC encounters from different ICMEs rather than repeated measurements of a single cloud, the systematic decrease of the fitted axial field strength $B_0$ with $\log(r)$ suggests that MCs undergo expansion during heliospheric propagation, leading to a progressive dilution of the axial magnetic field.

\subsection{Twist properties of MCs}\label{subsec:twist}

In the GH reconstruction, we characterize the field-line winding of an MC 
using the turn density, which measures how many turns a magnetic
field line makes around the flux-rope axis per unit axial length. Throughout
this work, all length-like quantities are first expressed in the time domain by
assuming that the solar-wind speed is approximately constant within each MC
interval, so that time can be used as a proxy for distance. Accordingly, we
denote by $\tau_t$ the turn density per unit time, reported in units of
turns per hour (turns~h$^{-1}$). In this paper, $\tau_t$ denotes turn density in the time domain (turns~h$^{-1}$), 
while $\tau \equiv \tau_{\rm AU}$ denotes turn density in the spatial domain (turns~AU$^{-1}$), unless noted otherwise.

Within the uniform-twist GH framework, the twist-related parameter $\omega$
links the rope scale to the local winding through
\begin{equation}
	\omega \equiv R_t\,T_t,
\end{equation}
where $R_t$ denotes the flux-rope cross-sectional scale (measured in hours in the
time domain) and $T_t$ is the local twist measure. In terms of the time-domain
turn density, $T_t = 2\pi \tau_t$, so that
\begin{equation}
	\tau_t = \frac{\omega}{2\pi R_t}.
\end{equation}
For a fixed $\omega$, the inferred turn density therefore scales as $R^{-1}$,
so an $R^{-1}$-type upper envelope in $\tau_t$ naturally corresponds to an
approximately constant upper bound in $\omega$.
\added{Thus, the upper envelope in the $\tau$--$R$ plane should be interpreted as a nearly constant upper boundary in $\omega$, rather than as a scale-independent upper limit on $\tau$ itself. For a given flux-rope radius, this boundary gives $\tau_{\max}(R)\simeq \omega_{\max}/(2\pi R)$.}
 
\begin{figure}[htbp]
	\centering
	\includegraphics[width=\textwidth]{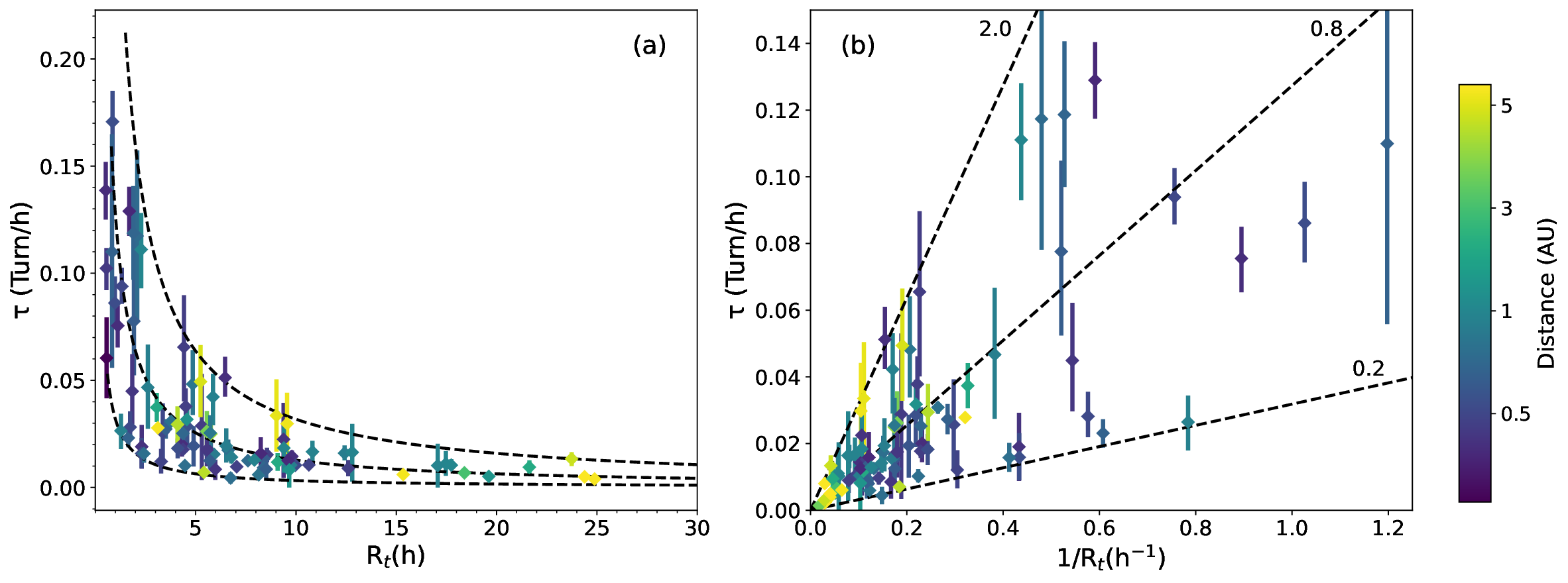}
\caption{
Scatterplots of the turn density $\tau$ versus $R$ (left) and $1/R$ (right). The color scale indicates the heliocentric distance of each MC event. The dashed lines denote constant-$\omega$ curves, $\tau=\omega/(2\pi R)$, for $\omega=2.0$, 0.8, and 0.2. The observed upper envelope is therefore expressed as an approximately constant boundary in $\omega$.
}

	\label{fig:tau_vs_R}
\end{figure}

 As shown in Figure~\ref{fig:tau_vs_R}, the turn densities populate
 a broad region in the $(\tau,R)$ plane, while the upper envelope decreases with
 increasing $R$ and is well approximated by an $R^{-1}$-type scaling. In particular,
 the envelope is closely tracked by $\omega=2$, corresponding to
 $\tau=1/(\pi R)$ in the GH framework. Thus, MCs with smaller inferred radii can
 reach much larger turn densities, whereas events with larger $R$ cluster below a
 nearly constant-$\omega$ boundary. This behavior is qualitatively consistent with
 the $\tau$--$R$ trend reported at 1~AU by \citet[][their Fig.~9]{Wang2016},
 \added{the observed upper envelope in the $\tau$--$R$ plane appears to depend on the global aspect ratio of the flux rope,} rather than following a single geometry-independent twist threshold.
A key reference for interpreting the observed upper envelope is the
Dungey--Loughhead (DL) geometric limit, which predicts that the critical
total twist angle scales with the aspect ratio as
\begin{equation}
	\Phi_{c}^{\rm DL} \approx 2\,\frac{l}{R},
\end{equation}
where $l$ is the axial length and $R$ is the rope radius
\citep{DungeyLoughhead1954,Bennett1999}. Using
$\Phi_T = 2\pi \tau_t l$, this limit can be rewritten as a constraint on the
turn density,
\begin{equation}
	\tau_{c}^{\rm DL} \approx \frac{1}{\pi R},
\end{equation}
which is equivalent, within the GH framework, to an upper bound on the model
parameter
\begin{equation}
	\omega \le \omega_{c}^{\rm DL} \equiv 2.
\end{equation}
In our multi-distance MC sample, almost no events lie above this DL boundary
, suggesting that it approximately traces the empirical upper envelope of the observed twist distribution of interplanetary flux ropes.

By contrast, the classical Hood--Priest threshold for a line-tied, uniformly
twisted GH rope ($\Phi_c \approx 2.5\pi$; \citealt{HoodPriest1981,Gold1960})
is exceeded by a substantial fraction of events. Because the HP criterion is derived for line-tied
coronal equilibria, while interplanetary MCs are not line-tied and evolve
geometrically during propagation, HP exceedance here is better interpreted as
evidence of configuration dependence rather than inconsistency. Overall, the DL
relation more closely follows the observed upper envelope over a broad radial range.

To express the turn density in physical units of turns per astronomical unit
(turns~AU$^{-1}$), we convert the time-domain turn density into a spatial one
by using the solar-wind speed as a space--time mapping along the flux-rope
axis. Within the uniform-twist GH framework, the turn density per unit
distance is given by
\begin{equation}
	\tau_{\rm AU}
	= \frac{\omega}{2\pi R}
	= \frac{\omega}{2\pi R_t\, v_{\rm sw}}
	= \frac{\tau_t}{v_{\rm sw}},
\end{equation}
where $R_t$ is the flux-rope scale in the time domain, $\tau_t$ is the turn
density in the unit of time, and $v_{\rm sw}$ is the solar-wind speed.

For MC events with available solar-wind speed measurements, we use the mean
solar-wind speed during the MC interval as the traversal speed. For MC events
without solar-wind speed data, we adopt a typical value of
$400~\mathrm{km\,s^{-1}}$ as the traversal speed (e.g., \citealt{Abbo2016,Zhuang2019}).

\begin{figure*}[ht!]
	\plotone{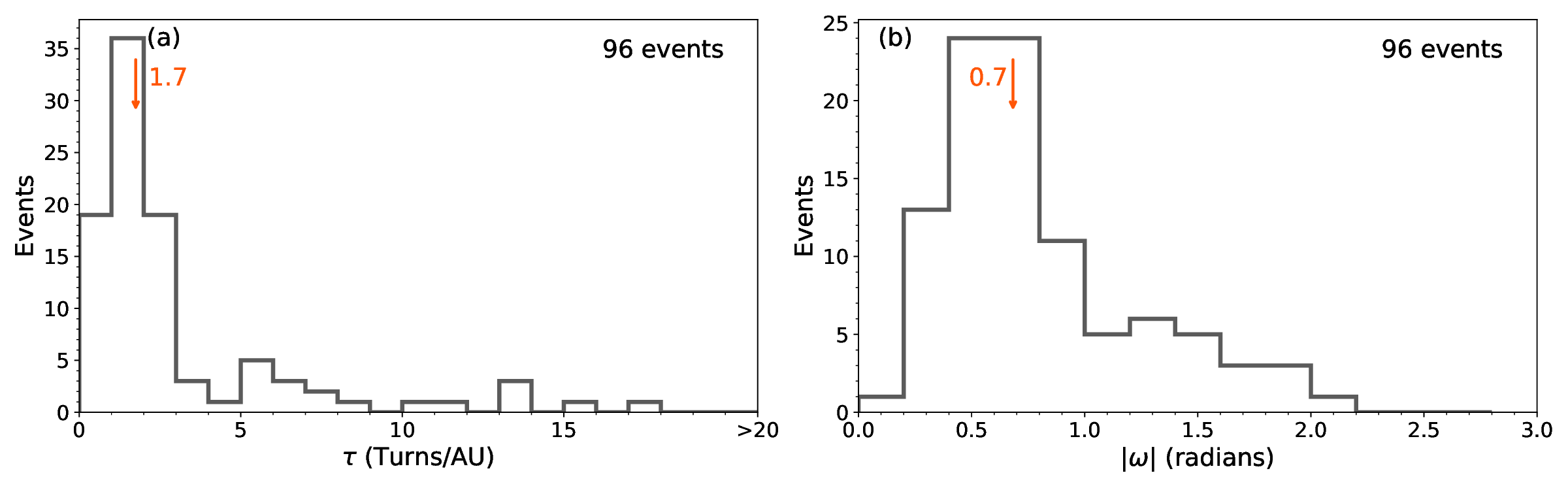}
	\caption{Histograms of the derived (a)~$\tau$ and (b)~$\omega$ obtained from all analyzed events. The arrows (orange) denote the median values in each panel. 
		\label{fig:tau_omega_distribution}}
\end{figure*}

Using this distance-based convention, we compute $\tau_{\rm AU}$ and the GH twist parameter $\omega$ for all 96 MC events and summarize their distributions in Figure~\ref{fig:tau_omega_distribution}. The distribution of $\tau_{\rm AU}$ has a median value of $1.7~{\rm turns~AU^{-1}}$, lower than the $3.6~{\rm turns~AU^{-1}}$ reported at 1 AU by \citet{Wang2016}. This difference may reflect the broader heliocentric coverage of the present sample, the adopted space--time conversion, and the empirical correction applied to GH-derived twist-related quantities. The corresponding distribution of $\omega$ has a median of 0.7, comparable to the representative value $\sim 0.6$ found by  \citet{Wang2016}. Together, these results indicate that interplanetary flux ropes commonly possess appreciable field-line winding, while their absolute turn-density values remain sensitive to the adopted normalization and sample selection.

\begin{figure*}[ht!]
	\plotone{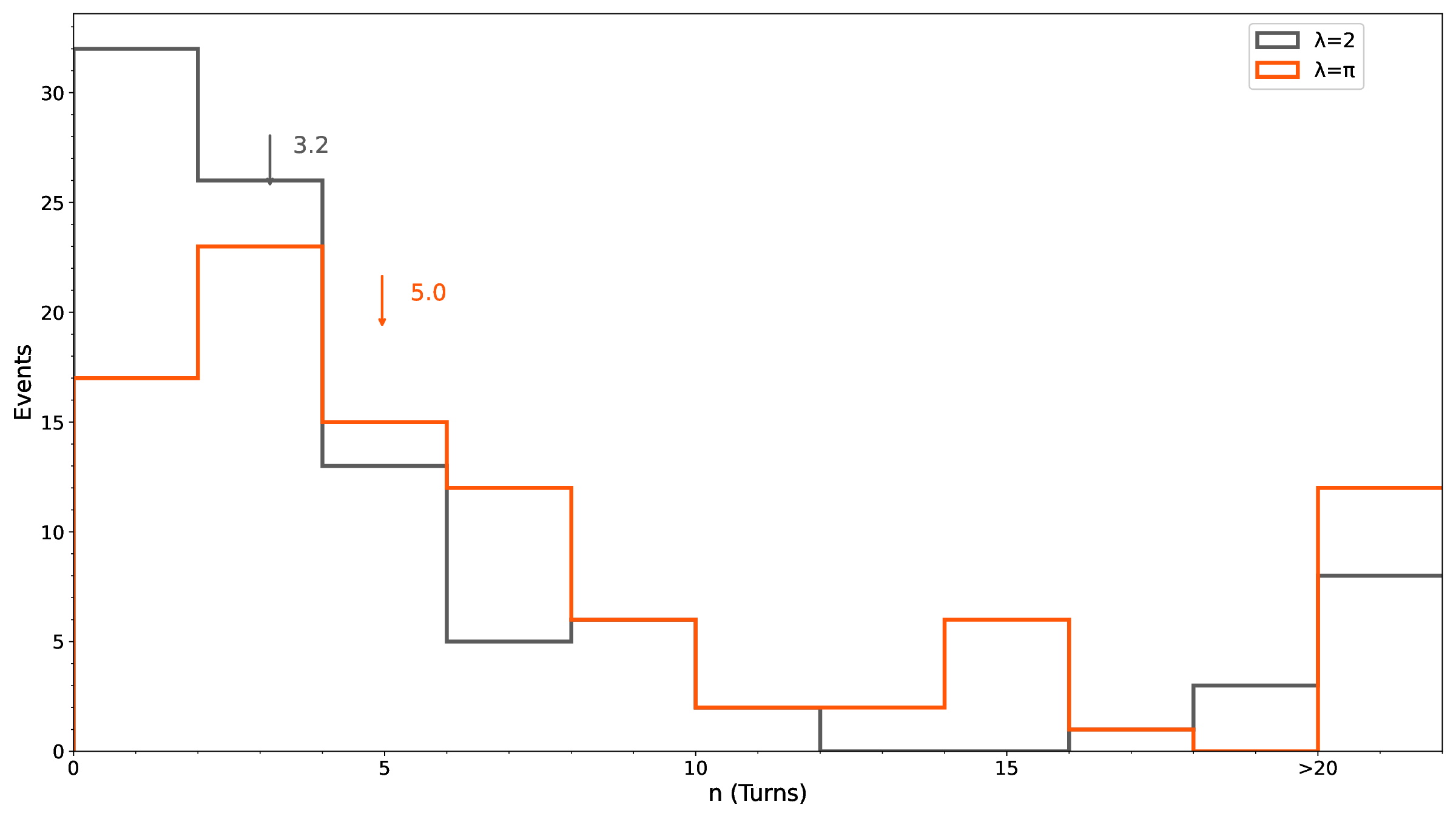}
	\caption{
		Histograms of the total number of magnetic field-line turns ($n$) derived for all MC events.
		We parameterize the assumed flux-rope axis length as $\lambda d$, where $d$ is the Sun--spacecraft distance at the observation time.
		The black curve uses $\lambda=2$, corresponding to an axis length of twice the Sun--spacecraft distance, while the orange curve uses $\lambda=\pi$, corresponding to a circular-arc length with $d$ taken as the diameter.
		The downward arrows mark the median values of 3.2 and 5.0 turns for the two assumptions, respectively.
	}
	\label{fig:tau}
\end{figure*}

Furthermore, after obtaining the distance-based turn density, we make a
rough estimate of the total number of field-line turns in each MC by
combining $\tau_{\rm AU}$ with simple geometric estimates of the flux-rope axial length.
For the flux-rope axial length $l$, we consider two simple prescriptions to bracket the uncertainty. Let $d$ be the heliocentric distance of the spacecraft at the observation time. We take (1) a minimal estimate $l_{\min}=2d$, and (2) a circular-arc estimate $l_{\rm circ}\approx \pi d$.
The total number of turns is then estimated as
\begin{equation}
	n \sim \tau_{\rm AU}\,l .
\end{equation}
Figure~\ref{fig:tau} summarizes the distribution of the estimated total field-line turns $n$ for our MC sample under two simple choices of the axial length $l$. Using the shorter length assumption (black histogram), the median is $n=3.2$, whereas adopting the longer arc-like length (orange histogram) yields a larger median of $n=5.0$ (downward arrows in the figure). Notably, under the shorter-length assumption, 80 out of the 96 MCs have twist numbers exceeding 1.25 turns, and under the longer arc-like length assumption, nearly all events (93/96) exceed 1.25 turns. Regardless of the length prescription, \added{these results indicate that interplanetary flux ropes are commonly multi-turn structures. However, the estimated values of $n$ depend explicitly on the assumed axial length, so comparisons with idealized twist thresholds should be interpreted cautiously.}

%When examining the heliospheric distribution of twist density and total twist (Figure~\ref{fig:tau_n_vs_r}), we find that $\tau$ generally decreases with increasing heliocentric distance. This behavior is consistent with progressive axial stretching of MC flux ropes during propagation, which dilutes the winding per unit length even if the global structure remains coherent. By contrast, the total number of turns $n$ does not show a systematic decrease with distance, but instead remains broadly scattered across the sampled radial range. 

\begin{figure*}[ht!]
	\plotone{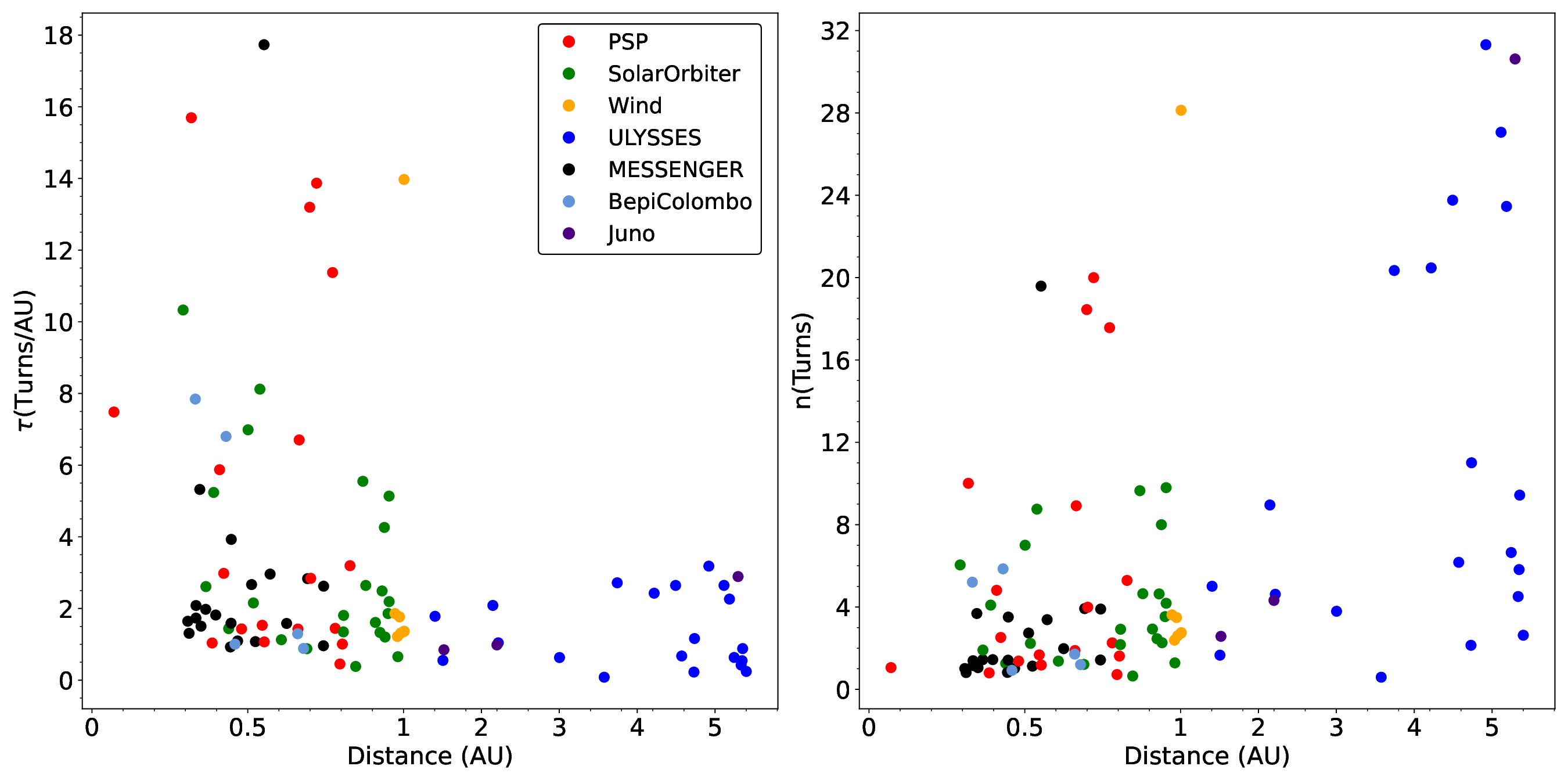}
\caption{Distributions of twist density and total twist number for the MC sample along the heliocentric distance in astronomical units (AU).
	Left panel: twist density $\tau$ versus heliocentric distance. Right panel: total twist number $n$ versus heliocentric distance.
	\label{fig:tau_n_vs_r}}
\end{figure*}

When examining the heliospheric distribution of twist density and total twist (Figure~\ref{fig:tau_n_vs_r}), we find that $\tau$ generally decreases with increasing heliocentric distance. This behavior is consistent with progressive axial stretching of MC flux ropes during propagation, which dilutes the winding per unit length even if the global structure remains coherent. \added{By contrast, the total number of turns $n$ remains broadly scattered across the sampled radial range and does not show a similarly clear radial organization. This is expected because $n$ is a global integral quantity and depends explicitly on the assumed axial length of the flux rope.}

% Notably,  when sparse data points in the $r\sim 2-4$ AU range are excluded, the twist number $n$ exhibits a tendency to increase with heliocentric distance. A plausible explanation for this behavior is that the magnetic twist is not uniformly distributed along the flux-rope axis. This  further illustrates that the evolution of twist in interplanetary flux ropes is  governed by more complex dynamic processes, rather than simple geometric expansion alone.

\section{Summary and Discussion}
\added{Using a uniform-twist GH reconstruction, we analyzed 96 MCs spanning a broad range of heliocentric distances and quantified their core field strength ($B_0$), turn density ($\tau$), GH parameter ($\omega$), and integrated turn number ($n$). The $\tau$--$R$ distribution shows an upper envelope with an approximate $R^{-1}$ scaling, which corresponds in the GH parametrization to a nearly constant upper boundary in $\omega=2\pi R\tau$, close to $\omega\sim2$. This result indicates that the winding density is constrained in a scale-dependent sense: for a given flux-rope radius, $\tau_{\max}(R)\simeq\omega_{\max}/(2\pi R)$. The observed envelope is consistent with a Dungey--Loughhead-type geometric scaling, but should not be interpreted as a scale-independent upper limit on $\tau$ itself or as a direct instability threshold for CME eruption.}

%Using a uniform-twist GH reconstruction, we analyzed 96 MCs spanning a broad range of heliocentric distances and quantified their core field strength ($B_0$), twist density ($\tau$), GH parameter ($\omega$), and integrated turn number ($n$). The main implication is that the observed upper-envelope behavior is more clearly organized by the local winding density  $\tau$ than by the integrated quantity $n$. The $\tau$--$R$ distribution shows a clear upper envelope with an approximate $R^{-1}$ scaling, equivalent in the GH parametrization to a nearly constant upper boundary in $\omega$ (close to $\omega\!\sim\!2$ for the envelope).  For comparison, the observed envelope can be expressed in a form similar to the Dungey–Loughhead geometric scaling. In its total-twist form,$\Phi_c^{\rm DL}\!\approx\!2l/R$, the threshold is explicitly aspect-ratio dependent; rewritten in terms of turn density, it gives $\tau_c^{\rm DL}\!\approx\!1/(\pi R)$, and in GH variables, $\omega_c^{\rm DL}\!=\!2$. The observed envelope follows a consistent geometric scaling of this form, indicating that the observed upper boundary appears to follow a geometric scaling associated with the global aspect ratio and the flux-rope cross-sectional scale. However, the present analysis is intended to characterize empirical trends in interplanetary flux-rope structure rather than to establish a direct instability criterion for CME eruption.

\citet{Wang2016} reported ICME observations at the L1 point, which are consistent 
with our results covering 0.07-5.40 AU. Their work also reveals a clear upper boundary 
in the $\tau$--$R$ relation, implying that such a boundary cannot be fully explained within 
a static framework and likely requires a dynamic theoretical model. Our findings 
therefore provide useful observational constraints for future dynamic modeling of ICMEs.

%\added{The estimated total turn number $n$ shows no similarly clear radial organization in the present sample. This is expected because $n$ is a global integral quantity and depends on the assumed axial length of the flux rope.}

The radial trends further support this picture. Both $B_0$ and $\tau$ decrease with
distance, consistent with expansion and axial stretching that dilute field strength
and winding density per unit length. \added{In contrast, the estimated total turn number $n$ remains broadly scattered and shows no similarly clear radial organization in the present sample. This behavior is expected because $n$ is a global integral quantity and depends on the adopted axial-length prescription. Overall, the $\tau$--$R$ distribution is more naturally organized by an approximately constant-$\omega$ envelope, whereas the interpretation of $n$ requires additional assumptions about the global flux-rope geometry.}

%However, $n$ does not exhibit a comparable, systematic upper-bound evolution with distance. Although the statistics remain limited in the  $r\sim 2-4$ AU range, $n$ gradually increases with radial distance, suggesting longitudinal nonuniformity in the twist distribution. Overall, $\tau$ acts as a local geometric diagnostic, while $n$ is a global integral quantity with weaker direct sensitivity to radial evolution.

These conclusions should be interpreted with several caveats, including limited
solar-wind speed availability for part of the sample and the dependence of $n$ on
the adopted axis-length prescriptions. Even with these uncertainties, the same
qualitative message persists:
the observed upper-envelope behavior of MC winding is more naturally organized by an aspect-ratio-controlled (DL-like) scaling than by a single geometry-independent twist threshold.

\section*{Acknowledgements}

We thank the \textit{PSP, Solar Orbiter, Wind, ULYSSES, MESSENGER, BepiColombo, Juno} mission team for providing the spacecraft data used in this study, and we are grateful to the anonymous referee for constructive feedback that improved the manuscript.
The model can be run and tested online at \url{http://space.ustc.edu.cn/dreams/mc_fitting/}.
This research was supported by the Chinese Academy of Sciences through the B-type Strategic Priority Program (XDB41000000), the Specialized Research Fund for State Key Laboratory of Solar Activity and Space Weather and the National Natural Science Foundation of China (Grant 42274203).
\added{B.S.-C. acknowledges support through STFC Ernest Rutherford Fellowship ST/V004115/1 and STFC grant ST/Y000439/1.}
			
\clearpage			
\begin{center}
	\begin{longtable}{c c c @{ } c c c c c c c c c c}
		\caption{Magnetic-cloud events and fitted GH-model parameters.\label{tab:long}}\\
		& \multicolumn{2}{c}{MC Interval} & \multicolumn{7}{c}{Modeled Parameters} \\
		\cline{2-3} \cline{5-11}
		NO. & $t_0$ & $\Delta t$ (h) &  & $B_0$ (nT) & $R$ (h) & $\theta$ ($^\circ$) & $\phi$ ($^\circ$) & $d$  & $\Delta t_c$ (h) & $\tau$ & $v_{\text{sw}}$ & $Q$ \\
		\hline
		\endfirsthead
        & \multicolumn{2}{c}{MC Interval} & \multicolumn{7}{c}{Modeled Parameters} \\
	\cline{2-3} \cline{5-11}
	NO. & $t_0$ & $\Delta t$ (h) &  & $B_0$ (nT) & $R$ (h) & $\theta$ ($^\circ$) & $\phi$ ($^\circ$) & $d$  & $\Delta t_c$ (h) & $\tau$ & $v_{\text{sw}}$ & $Q$ \\
	\hline
	\endhead		

%\cline{2-13} 
	\hline
\endfoot	

		\hline \hline
		\endlastfoot
1 &  1991-02-27 00:42:00  & 19.0 &         & 3.3 & 9.1 & -9.4 & 287.9 & -0.1 & 6.8 & $-0.030_{-0.010}^{+0.010}$   & 470.7 &  1 \\
2 &  1991-09-18 13:12:00  & 23.0 &         & 1.9 & 5.8 & 1.6 & 332.2 & -0.4 & 10.3 & $0.062_{-0.011}^{+0.010}$   & 427.7 &  2 \\
3 &  1991-11-18 09:29:00  & 72.0 &         & 1.5 & 34.8 & -19.8 & 252.2 & 0.1 & 18.5 & $-0.008_{-0.001}^{+0.002}$   & 564.0 &  1 \\
4 &  1992-06-22 05:04:00  & 62.0 &         & 0.8 & 24.4 & 6.7 & 310.0 & 0.2 & 27.9 & $-0.012_{-0.004}^{+0.004}$   & 379.7 &  2 \\
5 &  1995-02-03 07:58:00  & 24.0 &         & 6.0 & 4.6 & -20.2 & 186.6 & -0.3 & 14.1 & $0.079_{-0.012}^{+0.014}$   & 740.6 &  2 \\
6 &  1995-10-18 19:06:00  & 79.0 &         & 47.2 & 14.5 & -11.9 & 59.3 & 0.4 & 15.7 & $-0.046_{-0.014}^{+0.015}$   & 513.7 &  1 \\
7 &  1996-12-24 03:07:30  & 64.0 &         & 18.9 & 17.1 & 29.8 & 246.3 & -0.4 & 16.3 & $-0.026_{-0.005}^{+0.005}$   & 491.5 &  1 \\
8 &  1997-01-09 15:36:00  & 46.0 &         & 3.8 & 23.7 & 51.2 & 98.6 & -0.3 & 15.4 & $-0.033_{-0.008}^{+0.008}$   & 477.6 &  2 \\
9 &  1997-05-26 15:33:30  & 42.0 &         & 23.5 & 2.3 & -5.0 & 175.0 & -0.5 & 16.3 & $-0.278_{-0.030}^{+0.025}$   & 476.3 &  1 \\
10 &  1997-08-30 03:18:00  & 30.0 &         & 2.1 & 15.3 & -30.3 & 294.0 & -0.4 & 14.7 & $0.015_{-0.003}^{+0.003}$   & 397.0 &  1 \\
11 &  1997-11-13 05:07:00  & 54.0 &         & 1.0 & 24.9 & 21.9 & 295.0 & -0.1 & 31.9 & $0.010_{-0.004}^{+0.005}$   & 387.4 &  2 \\
12 &  1998-02-27 10:48:00  & 64.0 &         & 1.0 & 34.2 & -14.3 & 275.0 & -0.4 & 31.2 & $0.005_{-0.001}^{+0.001}$   & 369.8 &  1 \\
13 &  1998-08-20 11:25:30  & 43.0 &         & 17.8 & 17.7 & 4.7 & 85.9 & 0.1 & 15.0 & $-0.026_{-0.006}^{+0.006}$   & 418.5 &  1 \\
14 &  1998-08-28 11:45:00  & 72.0 &         & 2.5 & 33.7 & -7.6 & 239.2 & -0.4 & 35.2 & $-0.020_{-0.004}^{+0.004}$   & 378.2 &  1 \\
15 &  1999-09-14 13:49:00  & 11.0 &         & 2.1 & 5.4 & -14.1 & 301.5 & -0.5 & 4.6 & $-0.018_{-0.002}^{+0.002}$   & 433.5 &  1 \\
16 &  1999-10-07 04:14:00  & 15.0 &         & 3.9 & 4.1 & 11.6 & 152.7 & -0.4 & 7.2 & $-0.074_{-0.022}^{+0.021}$   & 462.1 &  2 \\
17 &  2000-03-31 12:16:00  & 20.0 &         & 8.4 & 5.6 & 7.2 & 147.7 & -0.2 & 9.8 & $-0.065_{-0.024}^{+0.024}$   & 400.3 &  1 \\
18 &  2000-08-10 11:25:00  & 47.0 &         & 1.8 & 18.4 & -33.2 & 142.3 & 0.3 & 23.1 & $0.017_{-0.005}^{+0.005}$   & 448.7 &  1 \\
19 &  2001-07-23 10:23:00  & 40.0 &         & 7.1 & 19.6 & -15.9 & 239.1 & -0.5 & 19.6 & $-0.013_{-0.002}^{+0.002}$   & 387.5 &  1 \\
20 &  2001-10-30 02:17:00  & 44.0 &         & 2.6 & 3.1 & -6.8 & 181.2 & 0.5 & 21.7 & $0.093_{-0.017}^{+0.017}$   & 744.4 &  2 \\
21 &  2002-06-14 08:24:00  & 110.0 &         & 2.0 & 54.7 & 7.5 & 82.5 & 0.1 & 32.6 & $0.003_{-0.001}^{+0.001}$   & 705.1 &  2 \\
22 &  2003-10-04 21:02:00  & 58.0 &         & 3.0 & 9.6 & 0.1 & 161.1 & -0.2 & 30.0 & $-0.074_{-0.036}^{+0.036}$   & 546.8 &  2 \\
23 &  2004-07-24 11:56:30  & 106.0 &         & 31.7 & 9.4 & -19.6 & 330.5 & 0.3 & 11.0 & $-0.046_{-0.005}^{+0.005}$   & 444.0 &  1 \\
24 &  2005-05-06 11:11:00  & 19.0 &         & 1.0 & 9.0 & -55.5 & 123.5 & 0.0 & 9.4 & $0.084_{-0.042}^{+0.042}$   & 526.9 &  2 \\
25 &  2005-08-03 08:30:00  & 17.0 &         & 9.6 & 5.2 & 1.4 & 33.7 & -0.4 & 9.0 & $0.123_{-0.041}^{+0.043}$   & 644.7 &  1 \\
26 &  2007-05-04 20:06:00  & 28.0 &         & 27.1 & 4.6 & 13.7 & 170.0 & 0.5 & 14.8 & $0.071_{-0.024}^{+0.027}$   & \nodata &  1 \\
27 &  2007-06-16 04:54:30  & 17.0 &         & 9.7 & 8.4 & 13.7 & 79.8 & 0.0 & 8.8 & $-0.023_{-0.005}^{+0.005}$   & \nodata &  1 \\
28 &  2007-06-16 22:19:30  & 10.4 &         & 15.5 & 4.4 & 45.6 & 320.8 & 0.0 & 5.5 & $-0.063_{-0.014}^{+0.014}$   & \nodata &  1 \\
29 &  2008-04-06 11:12:00  & 7.8 &         & 25.5 & 3.5 & -39.4 & 303.5 & -0.0 & 10.2 & $0.068_{-0.011}^{+0.011}$   & \nodata &  1 \\
30 &  2009-01-20 06:07:00  & 17.8 &         & 32.1 & 8.6 & -17.4 & 111.9 & 0.2 & 8.9 & $0.038_{-0.009}^{+0.008}$   & \nodata &  1 \\
31 &  2009-02-20 19:42:00  & 4.7 &         & 35.1 & 2.3 & -15.7 & 75.0 & -0.1 & 2.2 & $0.048_{-0.026}^{+0.025}$   & \nodata &  1 \\
32 &  2009-08-11 16:34:30  & 11.2 &         & 22.4 & 5.8 & -50.6 & 285.8 & 0.3 & 7.8 & $-0.038_{-0.016}^{+0.017}$   & \nodata &  1 \\
33 &  2009-09-23 14:04:00  & 17.0 &         & 51.6 & 9.8 & 0.0 & 270.1 & -0.5 & 6.0 & $-0.036_{-0.006}^{+0.007}$   & \nodata &  1 \\
34 &  2009-12-26 16:31:00  & 25.8 &         & 22.3 & 12.6 & 65.3 & 120.5 & 0.0 & 12.9 & $0.022_{-0.009}^{+0.010}$   & \nodata &  1 \\
35 &  2010-11-05 17:04:30  & 20.2 &         & 61.1 & 10.6 & -53.6 & 235.0 & -0.4 & 9.9 & $-0.026_{-0.003}^{+0.003}$   & \nodata &  1 \\
36 &  2010-12-09 01:52:30  & 12.6 &         & 46.9 & 4.3 & -27.1 & 214.9 & 0.0 & 6.2 & $0.050_{-0.014}^{+0.015}$   & \nodata &  1 \\
37 &  2011-01-05 08:20:30  & 19.7 &         & 27.5 & 10.0 & 19.6 & 265.0 & -0.2 & 9.8 & $-0.026_{-0.003}^{+0.004}$   & \nodata &  1 \\
38 &  2011-02-12 05:35:30  & 12.1 &         & 38.8 & 3.4 & -6.7 & 212.7 & -0.1 & 6.0 & $-0.064_{-0.034}^{+0.034}$   & \nodata &  1 \\
39 &  2011-03-09 04:28:00  & 10.4 &         & 87.5 & 5.7 & -21.4 & 76.4 & -0.5 & 2.8 & $0.042_{-0.012}^{+0.013}$   & \nodata &  1 \\
40 &  2011-11-14 10:43:00  & 14.5 &         & 44.5 & 5.6 & -24.0 & 45.0 & -0.1 & 5.0 & $0.044_{-0.019}^{+0.020}$   & \nodata &  2 \\
41 &  2011-11-23 10:42:00  & 15.1 &         & 126.0 & 6.5 & 55.1 & 341.1 & -0.2 & 4.5 & $0.128_{-0.022}^{+0.024}$   & \nodata &  2 \\
42 &  2012-05-25 11:20:00  & 27.0 &         & 80.5 & 9.6 & 29.3 & 144.3 & -0.1 & 15.0 & $-0.031_{-0.008}^{+0.008}$   & \nodata &  2 \\
43 &  2012-11-16 13:37:00  & 64.0 &         & 5.5 & 21.6 & 20.3 & 130.9 & -0.4 & 25.5 & $-0.024_{-0.000}^{+0.000}$   & 589.4 &  1 \\
44 &  2012-11-21 09:10:00  & 16.8 &         & 88.0 & 8.2 & -5.0 & 106.4 & 0.2 & 8.5 & $-0.040_{-0.018}^{+0.019}$   & \nodata &  2 \\
45 &  2013-05-02 10:34:00  & 108.0 &         & 6.6 & 9.7 & -13.3 & 115.0 & 0.4 & 9.7 & $0.020_{-0.003}^{+0.003}$   & 568.6 &  1 \\
46 &  2013-09-13 00:42:00  & 13.1 &         & 58.7 & 4.5 & 21.6 & 215.0 & -0.3 & 8.2 & $-0.095_{-0.021}^{+0.021}$   & \nodata &  2 \\
47 &  2014-04-11 06:55:30  & 68.0 &         & 15.1 & 17.5 & -50.8 & 139.7 & -0.3 & 19.0 & $-0.028_{-0.007}^{+0.006}$   & 417.3 &  1 \\
48 &  2016-01-17 00:09:00  & 132.0 &         & 4.6 & 3.1 & 11.7 & 190.0 & -0.6 & 9.5 & $-0.070_{-0.007}^{+0.006}$   & 454.8 &  1 \\
49 &  2016-10-13 06:27:00  & 107.0 &         & 40.4 & 12.8 & -16.8 & 225.0 & 0.4 & 15.2 & $0.041_{-0.012}^{+0.013}$   & 408.4 &  1 \\
50 &  2019-03-15 12:16:00  & 5.6 &         & 31.4 & 2.3 & -10.5 & 130.6 & -0.3 & 2.7 & $-0.040_{-0.017}^{+0.018}$   & 430.4 &  1 \\
51 &  2019-03-24 03:46:30  & 13.9 &         & 30.3 & 6.0 & 17.4 & 125.0 & 0.2 & 6.9 & $-0.021_{-0.012}^{+0.011}$   & 340.0 &  1 \\
52 &  2019-09-10 10:45:00  & 3.0 &         & 52.2 & 0.5 & -10.0 & 191.0 & -0.7 & 1.4 & $0.347_{-0.034}^{+0.033}$   & 367.2 &  1 \\
53 &  2020-04-19 09:02:00  & 24.3 &         & 20.4 & 6.6 & 3.1 & 211.7 & 0.3 & 12.1 & $0.043_{-0.011}^{+0.012}$   & \nodata &  2 \\
54 &  2020-04-21 05:12:30  & 7.5 &         & 8.1 & 3.8 & 0.4 & 110.0 & 0.4 & 3.8 & $-0.077_{-0.005}^{+0.004}$   & 401.4 &  1 \\
55 &  2020-06-23 06:27:30  & 9.8 &         & 15.9 & 3.3 & 17.2 & 322.2 & 0.2 & 5.6 & $0.030_{-0.014}^{+0.015}$   & 351.8 &  1 \\
56 &  2020-07-22 08:59:00  & 14.7 &         & 7.1 & 6.7 & 15.7 & 294.8 & 0.1 & 7.2 & $-0.011_{-0.006}^{+0.007}$   & 406.5 &  1 \\
57 &  2020-10-28 13:21:30  & 3.3 &         & 17.0 & 1.6 & 11.9 & 103.2 & 0.2 & 1.7 & $-0.058_{-0.011}^{+0.010}$   & 337.8 &  1 \\
58 &  2020-12-01 03:25:30  & 5.3 &         & 31.8 & 2.4 & 64.3 & 7.0 & -0.1 & 2.5 & $-0.039_{-0.011}^{+0.011}$   & 650.9 &  1 \\
59 &  2021-05-10 09:10:00  & 21.6 &         & 21.8 & 12.4 & 23.8 & 247.3 & 0.0 & 13.2 & $0.040_{-0.008}^{+0.009}$   & 500.4 &  2 \\
60 &  2021-05-27 20:21:30  & 14.2 &         & 10.6 & 6.8 & -5.0 & 291.0 & 0.2 & 7.0 & $-0.035_{-0.007}^{+0.007}$   & 316.8 &  1 \\
61 &  2021-06-12 09:32:00  & 15.9 &         & 41.9 & 2.1 & -5.0 & 346.5 & -0.3 & 7.7 & $0.293_{-0.098}^{+0.100}$   & 428.7 &  1 \\
62 &  2021-06-22 22:33:00  & 12.4 &         & 18.7 & 6.5 & 5.0 & 271.3 & 0.3 & 6.1 & $0.048_{-0.019}^{+0.021}$   & 323.1 &  1 \\
63 &  2021-07-01 11:16:30  & 16.5 &         & 18.2 & 5.9 & 0.9 & 145.2 & -0.6 & 8.2 & $-0.039_{-0.001}^{+0.001}$   & \nodata &  1 \\
64 &  2021-07-26 01:34:00  & 9.4 &         & 7.3 & 4.5 & -12.4 & 115.0 & 0.3 & 4.8 & $-0.025_{-0.005}^{+0.005}$   & 314.6 &  1 \\
65 &  2021-09-12 15:40:00  & 15.8 &         & 27.5 & 1.9 & -7.0 & 9.4 & -0.5 & 8.1 & $-0.297_{-0.054}^{+0.055}$   & 355.5 &  1 \\
66 &  2021-09-26 23:07:00  & 13.5 &         & 18.5 & 7.6 & 75.0 & 120.1 & -0.5 & 6.7 & $-0.031_{-0.004}^{+0.004}$   & 359.8 &  1 \\
67 &  2021-11-04 01:33:00  & 18.4 &         & 18.5 & 8.1 & 55.0 & 325.0 & 0.1 & 9.1 & $0.015_{-0.004}^{+0.004}$   & 644.9 &  1 \\
68 &  2021-11-09 18:45:00  & 10.0 &         & 96.0 & 5.3 & -7.0 & 246.3 & 0.5 & 4.9 & $0.072_{-0.063}^{+0.061}$   & 401.3 &  1 \\
69 &  2021-12-06 21:07:00  & 33.0 &         & 24.0 & 8.5 & 24.4 & 97.4 & -0.1 & 8.5 & $-0.021_{-0.006}^{+0.006}$   & 472.1 &  1 \\
70 &  2022-01-29 11:23:30  & 15.0 &         & 32.1 & 8.4 & 10.1 & 85.0 & 0.5 & 7.1 & $-0.034_{-0.007}^{+0.007}$   & 391.9 &  1 \\
71 &  2022-03-11 23:27:30  & 8.2 &         & 120.4 & 4.4 & 15.2 & 265.0 & 0.4 & 3.4 & $0.044_{-0.005}^{+0.004}$   & 511.7 &  1 \\
72 &  2022-04-13 07:34:00  & 62.0 &         & 41.4 & 5.7 & -7.0 & 307.7 & 0.5 & 6.1 & $0.031_{-0.000}^{+0.000}$   & 578.7 &  1 \\
73 &  2022-05-21 16:27:00  & 9.4 &         & 28.0 & 4.9 & 50.9 & 305.0 & -0.4 & 4.9 & $0.121_{-0.036}^{+0.040}$   & 361.4 &  1 \\
74 &  2022-05-23 02:47:30  & 29.1 &         & 16.2 & 5.7 & 5.5 & 202.1 & 0.2 & 13.4 & $0.064_{-0.021}^{+0.021}$   & \nodata &  1 \\
75 &  2022-06-02 11:38:30  & 2.3 &         & 615.2 & 0.6 & 10.4 & 332.9 & 0.1 & 1.4 & $0.151_{-0.047}^{+0.047}$   & 335.4 &  1 \\
76 &  2022-07-25 12:38:00  & 14.7 &         & 35.3 & 7.9 & -72.1 & 223.5 & 0.4 & 6.5 & $-0.032_{-0.002}^{+0.003}$   & 814.4 &  1 \\
77 &  2022-08-18 16:01:30  & 8.2 &         & 59.1 & 4.1 & -11.1 & 269.3 & 0.1 & 4.1 & $-0.046_{-0.012}^{+0.012}$   & 710.3 &  1 \\
78 &  2022-10-02 22:56:30  & 18.6 &         & 59.3 & 9.4 & -60.1 & 270.8 & -0.2 & 8.7 & $-0.056_{-0.045}^{+0.043}$   & 357.7 &  2 \\
79 &  2023-01-05 02:22:00  & 104.0 &         & 27.4 & 10.8 & 15.9 & 141.2 & 0.3 & 3.6 & $0.042_{-0.005}^{+0.005}$   & 643.9 &  1 \\
80 &  2023-01-08 17:59:00  & 12.2 &         & 24.8 & 5.9 & 18.1 & 51.3 & 0.5 & 6.0 & $-0.106_{-0.033}^{+0.027}$   & 341.7 &  1 \\
81 &  2023-01-12 01:16:00  & 5.8 &         & 22.8 & 0.8 & -15.9 & 5.6 & -0.0 & 3.3 & $-0.275_{-0.135}^{+0.137}$   & 346.2 &  1 \\
82 &  2023-01-12 11:39:00  & 8.7 &         & 10.8 & 1.3 & 6.7 & 164.8 & 0.3 & 4.2 & $0.066_{-0.021}^{+0.020}$   & 501.5 &  1 \\
83 &  2023-02-19 00:52:00  & 29.0 &         & 79.4 & 1.1 & 10.2 & 176.9 & -0.0 & 6.0 & $-0.189_{-0.080}^{+0.082}$   & 485.5 &  1 \\
84 &  2023-03-21 13:28:30  & 6.4 &         & 82.7 & 1.3 & -15.8 & 356.2 & 0.7 & 3.2 & $-0.235_{-0.020}^{+0.022}$   & 558.4 &  1 \\
85 &  2023-04-10 10:26:00  & 3.1 &         & 210.6 & 1.7 & -30.1 & 284.5 & -0.5 & 1.3 & $0.322_{-0.029}^{+0.029}$   & 518.8 &  2 \\
86 &  2023-04-22 09:13:30  & 3.4 &         & 33.7 & 1.8 & -27.2 & 292.3 & 0.5 & 1.4 & $0.112_{-0.038}^{+0.043}$   & 356.3 &  2 \\
87 &  2023-05-03 10:29:00  & 5.5 &         & 53.0 & 1.0 & 0.0 & 15.0 & -0.7 & 2.9 & $0.215_{-0.030}^{+0.031}$   & 440.6 &  1 \\
88 &  2023-06-27 03:33:30  & 20.8 &         & 19.9 & 9.7 & 33.5 & 125.0 & -0.3 & 6.3 & $-0.026_{-0.002}^{+0.002}$   & 360.5 &  1 \\
89 &  2023-07-19 18:01:00  & 2.9 &         & 25.8 & 1.9 & 40.2 & 65.1 & -0.7 & 1.5 & $0.194_{-0.063}^{+0.068}$   & 480.9 &  2 \\
90 &  2023-07-22 20:37:00  & 5.5 &         & 14.2 & 2.6 & -13.6 & 246.1 & -0.3 & 2.8 & $-0.117_{-0.048}^{+0.050}$   & 455.3 &  2 \\
91 &  2023-09-08 22:21:30  & 9.8 &         & 32.1 & 4.9 & 5.0 & 98.2 & -0.2 & 3.5 & $-0.048_{-0.024}^{+0.023}$   & 713.3 &  1 \\
92 &  2023-09-17 05:31:30  & 8.8 &         & 66.9 & 0.6 & 4.4 & 5.1 & -0.3 & 3.9 & $-0.256_{-0.025}^{+0.024}$   & 723.8 &  1 \\
93 &  2023-09-22 03:46:00  & 95.0 &         & 39.7 & 7.0 & 11.4 & 239.1 & -0.1 & 8.1 & $-0.024_{-0.007}^{+0.007}$   & 685.6 &  1 \\
94 &  2023-10-28 14:13:30  & 6.7 &         & 19.1 & 1.7 & 18.8 & 203.4 & 0.3 & 3.3 & $0.070_{-0.016}^{+0.018}$   & 542.7 &  1 \\
95 &  2023-11-12 03:59:00  & 16.4 &         & 16.1 & 8.3 & 54.9 & 297.4 & -0.3 & 8.1 & $0.020_{-0.004}^{+0.003}$   & 377.5 &  1 \\
96 &  2024-02-23 06:42:00  & 34.0 &         & 79.5 & 4.4 & -17.1 & 170.0 & -0.1 & 12.9 & $-0.164_{-0.065}^{+0.064}$   & 611.6 &  2 \\
	\end{longtable}
\end{center}
\clearpage		

\bibliography{sample701}{}

@ARTICLE{Abbo2016,
	author = {{Abbo}, L. and {Ofman}, L. and {Antiochos}, S.~K. and {Hansteen}, V.~H. and {Harra}, L. and {Ko}, Y.-K. and {Lapenta}, G. and {Li}, B. and {Riley}, P. and {Strachan}, L. and {von Steiger}, R. and {Wang}, Y.-M.},
	title = "{Slow Solar Wind: Observations and Modeling}",
	journal = {\ssr},
	year = 2016,
	month = nov,
	volume = {201},
	number = {1-4},
	pages = {55-108},
	doi = {10.1007/s11214-016-0264-1},
	adsurl = {https://ui.adsabs.harvard.edu/abs/2016SSRv..201...55A}
}

@ARTICLE{Bennett1999,
	author = {{Bennett}, K. and {Roberts}, B. and {Narain}, U.},
	title = "{Waves in Twisted Magnetic Flux Tubes}",
	journal = {\solphys},
	year = 1999,
	month = mar,
	volume = {185},
	number = {1},
	pages = {41-59},
	doi = {10.1023/A:1005141432432},
	adsurl = {https://ui.adsabs.harvard.edu/abs/1999SoPh..185...41B}
}

@article{Burlaga1981,
	author = {{Burlaga}, L. and {Sittler}, E. and {Mariani}, F. and {Schwenn}, R.},
	title = "{Magnetic loop behind an interplanetary shock: Voyager, Helios, and IMP 8 observations}",
	journal = {\jgr},
	year = 1981,
	month = aug,
	volume = {86},
	number = {A8},
	pages = {6673-6684},
	doi = {10.1029/JA086iA08p06673},
	adsurl = {https://ui.adsabs.harvard.edu/abs/1981JGR....86.6673B}
}

@ARTICLE{DungeyLoughhead1954,
	author = {{Dungey}, J.~W. and {Loughhead}, R.~E.},
	title = "{Twisted Magnetic Fields in Conducting Fluids}",
	journal = {Australian Journal of Physics},
	year = 1954,
	month = mar,
	volume = {7},
	pages = {5},
	doi = {10.1071/PH540005},
	adsurl = {https://ui.adsabs.harvard.edu/abs/1954AuJPh...7....5D}
}

@article{Gary2004,
author = {{Gary}, G. Allen and {Moore}, R.~L.},
title = "{Eruption of a Multiple-Turn Helical Magnetic Flux Tube in a Large Flare: Evidence for External and Internal Reconnection That Fits the Breakout Model of Solar Magnetic Eruptions}",
journal = {\apj},
year = 2004,
month = aug,
volume = {611},
number = {1},
pages = {545-556},
doi = {10.1086/422132},
adsurl = {https://ui.adsabs.harvard.edu/abs/2004ApJ...611..545G}
}

@article{Gold1960,
author = {{Gold}, T. and {Hoyle}, F.},
title = "{On the origin of solar flares}",
journal = {\mnras},
year = 1960,
month = jan,
volume = {120},
pages = {89},
doi = {10.1093/mnras/120.2.89},
adsurl = {https://ui.adsabs.harvard.edu/abs/1960MNRAS.120...89G}
}

@article{Guo2010,
author = {{Guo}, Y. and {Schmieder}, B. and {D{\'e}moulin}, P. and {Wiegelmann}, T. and {Aulanier}, G. and {T{\"o}r{\"o}k}, T. and {Bommier}, V.},
title = "{Coexisting Flux Rope and Dipped Arcade Sections Along One Solar Filament}",
journal = {\apj},
year = 2010,
month = may,
volume = {714},
number = {1},
pages = {343-354},
doi = {10.1088/0004-637X/714/1/343},
adsurl = {https://ui.adsabs.harvard.edu/abs/2010ApJ...714..343G}
}

@article{HoodPriest1981,
author = {{Hood}, A.~W. and {Priest}, E.~R.},
title = "{Critical conditions for magnetic instabilities in force-free coronal loops}",
journal = {Geophysical and Astrophysical Fluid Dynamics},
year = 1981,
month = jan,
volume = {17},
number = {1},
pages = {297-318},
doi = {10.1080/03091928108243687},
adsurl = {https://ui.adsabs.harvard.edu/abs/1981GApFD..17..297H}
}

@article{Inoue2011,
author = {{Inoue}, S. and {Kusano}, K. and {Magara}, T. and {Shiota}, D. and {Yamamoto}, T.~T.},
title = "{Twist and Connectivity of Magnetic Field Lines in the Solar Active Region NOAA 10930}",
journal = {\apj},
year = 2011,
month = sep,
volume = {738},
number = {2},
eid = {161},
pages = {161},
doi = {10.1088/0004-637X/738/2/161},
adsurl = {https://ui.adsabs.harvard.edu/abs/2011ApJ...738..161I}
}

@article{Inoue2012,
author = {{Inoue}, S. and {Shiota}, D. and {Yamamoto}, T.~T. and {Pandey}, V.~S. and {Magara}, T. and {Choe}, G.~S.},
title = "{Buildup and Release of Magnetic Twist during the X3.4 Solar Flare of 2006 December 13}",
journal = {\apj},
year = 2012,
month = nov,
volume = {760},
number = {1},
eid = {17},
pages = {17},
doi = {10.1088/0004-637X/760/1/17},
archivePrefix = {arXiv},
eprint = {1209.6131},
primaryClass = {astro-ph.SR},
adsurl = {https://ui.adsabs.harvard.edu/abs/2012ApJ...760...17I}
}

@article{KleinBurlaga1982,
author = {{Klein}, L.~W. and {Burlaga}, L.~F.},
title = "{Interplanetary magnetic clouds at 1 AU}",
journal = {\jgr},
year = 1982,
month = feb,
volume = {87},
number = {A2},
pages = {613-624},
doi = {10.1029/JA087iA02p00613},
adsurl = {https://ui.adsabs.harvard.edu/abs/1982JGR....87..613K}
}

@ARTICLE{Kruskal1958,
	author = {{Kruskal}, M.~D. and {Johnson}, J.~L. and {Gottlieb}, M.~B. and {Goldman}, L.~M.},
	title = "{Hydromagnetic Instability in a Stellarator}",
	journal = {Physics of Fluids},
	year = 1958,
	month = sep,
	volume = {1},
	number = {5},
	pages = {421-429},
	doi = {10.1063/1.1724359},
	adsurl = {https://ui.adsabs.harvard.edu/abs/1958PhFl....1..421K}
}

@ARTICLE{Liu2014,
	author = {{Liu}, Jiajia and {Wang}, Yuming and {Liu}, Rui and {Zhang}, Quanhao and {Liu}, Kai and {Shen}, Chenglong and {Wang}, S.},
	title = "{When and how does a Prominence-like Jet Gain Kinetic Energy?}",
	journal = {\apj},
	year = 2014,
	month = feb,
	volume = {782},
	number = {2},
	eid = {94},
	pages = {94},
	doi = {10.1088/0004-637X/782/2/94},
	archivePrefix = {arXiv},
	eprint = {1401.0736},
	primaryClass = {astro-ph.SR},
	adsurl = {https://ui.adsabs.harvard.edu/abs/2014ApJ...782...94L}
}

@article{Lundquist1950, 
	title={Magneto-hydrostatic fields},
	author={Lundquist, Stig},
	journal={Ark. Fys.},
	volume={2},
	pages={361--365},
	year={1950}
}

@article{Mostl2017,
author = {{M{\"o}stl}, C. and {Isavnin}, A. and {Boakes}, P.~D. and {Kilpua}, E.~K.~J. and {Davies}, J.~A. and {Harrison}, R.~A. and {Barnes}, D. and {Krupar}, V. and {Eastwood}, J.~P. and {Good}, S.~W. and {Forsyth}, R.~J. and {Bothmer}, V. and {Reiss}, M.~A. and {Amerstorfer}, T. and {Winslow}, R.~M. and {Anderson}, B.~J. and {Philpott}, L.~C. and {Rodriguez}, L. and {Rouillard}, A.~P. and {Gallagher}, P. and {Nieves-Chinchilla}, T. and {Zhang}, T.~L.},
title = "{Modeling observations of solar coronal mass ejections with heliospheric imagers verified with the Heliophysics System Observatory}",
journal = {Space Weather},
year = 2017,
month = jul,
volume = {15},
number = {7},
pages = {955-970},
doi = {10.1002/2017SW001614},
archivePrefix = {arXiv},
eprint = {1703.00705},
primaryClass = {astro-ph.SR},
adsurl = {https://ui.adsabs.harvard.edu/abs/2017SpWea..15..955M}
}

@article{Mostl2020,
author = {{M{\"o}stl}, Christian and {Weiss}, Andreas J. and {Bailey}, Rachel L. and {Reiss}, Martin A. and {Amerstorfer}, Tanja and {Hinterreiter}, J{\"u}rgen and {Bauer}, Maike and {McIntosh}, Scott W. and {Lugaz}, No{\'e} and {Stansby}, David},
title = "{Prediction of the In Situ Coronal Mass Ejection Rate for Solar Cycle 25: Implications for Parker Solar Probe In Situ Observations}",
journal = {\apj},
year = 2020,
month = nov,
volume = {903},
number = {2},
eid = {92},
pages = {92},
doi = {10.3847/1538-4357/abb9a1},
archivePrefix = {arXiv},
eprint = {2007.14743},
primaryClass = {astro-ph.SR},
adsurl = {https://ui.adsabs.harvard.edu/abs/2020ApJ...903...92M}
}

@ARTICLE{Pulkkinen2007,
	author = {{Pulkkinen}, Tuija},
	title = "{Space Weather: Terrestrial Perspective}",
	journal = {Living Reviews in Solar Physics},
	year = 2007,
	month = dec,
	volume = {4},
	number = {1},
	eid = {1},
	pages = {1},
	doi = {10.12942/lrsp-2007-1},
	adsurl = {https://ui.adsabs.harvard.edu/abs/2007LRSP....4....1P}
}

@article{Regnier2002,
author = {{R{\'e}gnier}, S. and {Amari}, T. and {Kersal{\'e}}, E.},
title = "{3D Coronal magnetic field from vector magnetograms: non-constant-alpha force-free configuration of the active region NOAA 8151}",
journal = {\aap},
year = 2002,
month = sep,
volume = {392},
pages = {1119-1127},
doi = {10.1051/0004-6361:20020993},
adsurl = {https://ui.adsabs.harvard.edu/abs/2002A&A...392.1119R}
}

@article{Romano2003,
author = {{Romano}, P. and {Contarino}, L. and {Zuccarello}, F.},
title = "{Eruption of a helically twisted prominence}",
journal = {\solphys},
year = 2003,
month = jun,
volume = {214},
number = {2},
pages = {313-323},
doi = {10.1023/A:1024257603143},
adsurl = {https://ui.adsabs.harvard.edu/abs/2003SoPh..214..313R}
}

@article{Shafranov1957,
	title={The structure of shock waves in a plasma},
	author={Shafranov, VD},
	journal={Soviet Phys. JETP},
	volume={5},
	year={1957},
	publisher={Academy of Sciences, USSR}
}

@ARTICLE{Schmidt2025,
	author = {{Schmidt}, Dirk and {Schad}, Thomas A. and {Yurchyshyn}, Vasyl and {Gorceix}, Nicolas and {Rimmele}, Thomas R. and {Goode}, Philip R.},
	title = "{Observations of fine coronal structures with high-order solar adaptive optics}",
	journal = {Nature Astronomy},
	year = 2025,
	month = aug,
	volume = {9},
	pages = {1148-1157},
	doi = {10.1038/s41550-025-02564-0},
	adsurl = {https://ui.adsabs.harvard.edu/abs/2025NatAs...9.1148S}
}

@ARTICLE{Schwenn2006,
	author = {{Schwenn}, Rainer},
	title = "{Space Weather: The Solar Perspective}",
	journal = {Living Reviews in Solar Physics},
	year = 2006,
	month = dec,
	volume = {3},
	number = {1},
	eid = {2},
	pages = {2},
	doi = {10.12942/lrsp-2006-2},
	adsurl = {https://ui.adsabs.harvard.edu/abs/2006LRSP....3....2S}
}

@article{Srivastava2010,
author = {{Srivastava}, A.~K. and {Zaqarashvili}, T.~V. and {Kumar}, Pankaj and {Khodachenko}, M.~L.},
title = "{Observation of Kink Instability During Small B5.0 Solar Flare on 2007 June 4}",
journal = {\apj},
year = 2010,
month = may,
volume = {715},
number = {1},
pages = {292-299},
doi = {10.1088/0004-637X/715/1/292},
archivePrefix = {arXiv},
eprint = {1004.1454},
primaryClass = {astro-ph.SR},
adsurl = {https://ui.adsabs.harvard.edu/abs/2010ApJ...715..292S}
}

@article{Vrsnak1991,
author = {{Vrsnak}, B. and {Ruzdjak}, V. and {Rompolt}, B.},
title = "{Stability of Prominences Exposing Helical like Patterns}",
journal = {\solphys},
year = 1991,
month = nov,
volume = {136},
number = {1},
pages = {151-167},
doi = {10.1007/BF00151701},
adsurl = {https://ui.adsabs.harvard.edu/abs/1991SoPh..136..151V}
}

@ARTICLE{Wang2015,
	author = {{Wang}, Yuming and {Zhou}, Zhenjun and {Shen}, Chenglong and {Liu}, Rui and {Wang}, S.},
	title = "{Investigating plasma motion of magnetic clouds at 1 AU through a velocity-modified cylindrical force-free flux rope model}",
	journal = {Journal of Geophysical Research (Space Physics)},
	year = 2015,
	month = mar,
	volume = {120},
	number = {3},
	pages = {1543-1565},
	doi = {10.1002/2014JA020494},
	archivePrefix = {arXiv},
	eprint = {1502.05112},
	primaryClass = {astro-ph.SR},
	adsurl = {https://ui.adsabs.harvard.edu/abs/2015JGRA..120.1543W}
}

@article{Wang2016,
author = {{Wang}, Yuming and {Zhuang}, Bin and {Hu}, Qiang and {Liu}, Rui and {Shen}, Chenglong and {Chi}, Yutian},
title = "{On the twists of interplanetary magnetic flux ropes observed at 1 AU}",
journal = {Journal of Geophysical Research (Space Physics)},
year = 2016,
month = oct,
volume = {121},
number = {10},
pages = {9316-9339},
doi = {10.1002/2016JA023075},
archivePrefix = {arXiv},
eprint = {1608.05607},
primaryClass = {astro-ph.SR},
adsurl = {https://ui.adsabs.harvard.edu/abs/2016JGRA..121.9316W}
}

@ARTICLE{Wang2017,
	author = {{Wang}, Wensi and {Liu}, Rui and {Wang}, Yuming and {Hu}, Qiang and {Shen}, Chenglong and {Jiang}, Chaowei and {Zhu}, Chunming},
	title = "{Buildup of a highly twisted magnetic flux rope during a solar eruption}",
	journal = {Nature Communications},
	year = 2017,
	month = nov,
	volume = {8},
	eid = {1330},
	pages = {1330},
	doi = {10.1038/s41467-017-01207-x},
	adsurl = {https://ui.adsabs.harvard.edu/abs/2017NatCo...8.1330W}
}

@ARTICLE{Wang2018,
author = {{Wang}, Yuming and {Shen}, Chenglong and {Liu}, Rui and {Liu}, Jiajia and {Guo}, Jingnan and {Li}, Xiaolei and {Xu}, Mengjiao and {Hu}, Qiang and {Zhang}, Tielong},
title = {Understanding the Twist Distribution Inside Magnetic Flux Ropes by Anatomizing an Interplanetary Magnetic Cloud},
journal = {Journal of Geophysical Research: Space Physics},
volume = {123},
number = {5},
pages = {3238-3261},
doi = {https://doi.org/10.1002/2017JA024971},
url = {https://agupubs.onlinelibrary.wiley.com/doi/abs/10.1002/2017JA024971},
eprint = {https://agupubs.onlinelibrary.wiley.com/doi/pdf/10.1002/2017JA024971},
year = {2018}
}

@article{Yan2001,
author = {{Yan}, Yihua and {Deng}, Yuanyong and {Karlick{\'y}}, Marian and {Fu}, Qijun and {Wang}, Shujuan and {Liu}, Yuying},
title = "{The Magnetic Rope Structure and Associated Energetic Processes in the 2000 July 14 Solar Flare}",
journal = {\apjl},
year = 2001,
month = apr,
volume = {551},
number = {1},
pages = {L115-L119},
doi = {10.1086/319829},
adsurl = {https://ui.adsabs.harvard.edu/abs/2001ApJ...551L.115Y}
}

@ARTICLE{Zhou2019,
	author = {{Zhou}, Zhenjun and {Cheng}, Xin and {Zhang}, Jie and {Wang}, Yuming and {Wang}, Dong and {Liu}, Lijuan and {Zhuang}, Bin and {Cui}, Jun},
	title = "{Why Do Torus-unstable Solar Filaments Experience Failed Eruptions?}",
	journal = {\apjl},
	year = 2019,
	month = jun,
	volume = {877},
	number = {2},
	eid = {L28},
	pages = {L28},
	doi = {10.3847/2041-8213/ab21cb},
	archivePrefix = {arXiv},
	eprint = {1905.00224},
	primaryClass = {astro-ph.SR},
	adsurl = {https://ui.adsabs.harvard.edu/abs/2019ApJ...877L..28Z}
}

@ARTICLE{Zhou2020,
	author = {{Zhou}, Zhenjun and {Liu}, Rui and {Cheng}, Xing and {Jiang}, Chaowei and {Wang}, Yuming and {Liu}, Lijuan and {Cui}, Jun},
	title = "{The Relationship between Chirality, Sense of Rotation, and Hemispheric Preference of Solar Eruptive Filaments}",
	journal = {\apj},
	year = 2020,
	month = mar,
	volume = {891},
	number = {2},
	eid = {180},
	pages = {180},
	doi = {10.3847/1538-4357/ab7666},
	archivePrefix = {arXiv},
	eprint = {2002.05007},
	primaryClass = {astro-ph.SR},
	adsurl = {https://ui.adsabs.harvard.edu/abs/2020ApJ...891..180Z}
}

@ARTICLE{Zhou2022,
	author = {{Zhou}, Zhenjun and {Jiang}, Chaowei and {Liu}, Rui and {Wang}, Yuming and {Liu}, Lijuan and {Cui}, Jun},
	title = "{The Rotation of Magnetic Flux Ropes Formed during Solar Eruption}",
	journal = {\apjl},
	year = 2022,
	month = mar,
	volume = {927},
	number = {1},
	eid = {L14},
	pages = {L14},
	doi = {10.3847/2041-8213/ac5740},
	archivePrefix = {arXiv},
	eprint = {2202.09073},
	primaryClass = {astro-ph.SR},
	adsurl = {https://ui.adsabs.harvard.edu/abs/2022ApJ...927L..14Z}
}

@article{Zhou2025,
author = {{Zhou}, Guiping},
title = "{Unveiling coronal fine structures with adaptive optics}",
journal = {Nature Astronomy},
year = 2025,
month = aug,
volume = {9},
pages = {1105-1106},
doi = {10.1038/s41550-025-02626-3},
adsurl = {https://ui.adsabs.harvard.edu/abs/2025NatAs...9.1105Z}
}

@article{Zhuang2019,
author = {{Zhuang}, Bin and {Wang}, Yuming and {Hu}, Youqiu and {Shen}, Chenglong and {Liu}, Rui and {Gou}, Tingyu and {Zhang}, Quanhao and {Li}, Xiaolei},
title = "{Numerical Simulations on the Deflection of Coronal Mass Ejections in the Interplanetary Space}",
journal = {\apj},
year = 2019,
month = may,
volume = {876},
number = {1},
eid = {73},
pages = {73},
doi = {10.3847/1538-4357/ab139e},
adsurl = {https://ui.adsabs.harvard.edu/abs/2019ApJ...876...73Z}
}
\bibliographystyle{aasjournalv7}

%\end{CJK*}
\end{document}